\documentclass[acmtog]{acmart}

\usepackage{xcolor}
\usepackage[noend]{algorithmic}
\usepackage{algorithm}
\usepackage{float}
\usepackage{outlines}
\usepackage{enumitem}
\usepackage[utf8]{inputenc}
\usepackage{wrapfig}
\usepackage{floatflt}
\AtBeginDocument{%
  \providecommand\BibTeX{{%
    \normalfont B\kern-0.5em{\scshape i\kern-0.25em b}\kern-0.8em\TeX}}}

\acmSubmissionID{261}

\begin{document}

\title{Technical Companion to Example-Based Procedural Modeling Using Graph Grammars}

\newcommand{\bN}{{\wedge}}
\newcommand{\bV}{{\vee}}
\newcommand{\hN}{{\cap}}
\newcommand{\hV}{{\cup}}
\newcommand{\br}[1]{\overline{#1}}

\author{Paul Merrell}
\email{paul@merrells.org}
\affiliation{%
\institution{ - }
\city{Redwood City}
\state{CA}
\country{USA}
}

\begin{abstract}

This is a companion piece to my paper on ``Example-Based Procedural Modeling 
Using Graph Grammars.'' This paper examines some of the theoretical issues in 
more detail. This paper discusses some more complex parts of the 
implementation, why certain algorithmic decisions were made, proves the 
algorithm can solve certain classes of problems, and examines other 
interesting theoretical questions.

\end{abstract}



\maketitle

\section{Introduction}

This is a companion piece to my paper on ``Example-Based Procedural Modeling 
Using Graph Grammars'' \cite{siggraph}. That paper raises many interesting theoretical 
questions. This paper will examine some of these issues in more detail. This 
paper will explain some more complex parts of the implementation, will 
explain why certain algorithmic decisions were made and their alternatives, 
will prove the algorithm can solve certain classes of problems, and examines 
other interesting theoretical questions.

I will assume the reader is familiar with the original paper \cite{siggraph}. Please refer
to that paper to understand motivation, related work, results, etc. I have also
written another supplemental paper. It shows additional results and
the grammars in more detail while this paper is targeted towards theoretic
issues.

\section{Graph Hierarchy Without Duplicates}
\label{Removing Duplicates}

A large graph is assembled through a series of gluing operations. Changing the order
of these operations does not change the result. Suppose that a grandparent $\br{a}x\br{x} \bN$
has two children $\br{a}\bN$ and $\br{ay}x\br{x} \bN$ and they share a child $\br{a}\br{y} \bN$:

\begin{figure}[H]
\centering
\includegraphics[width=8cm]{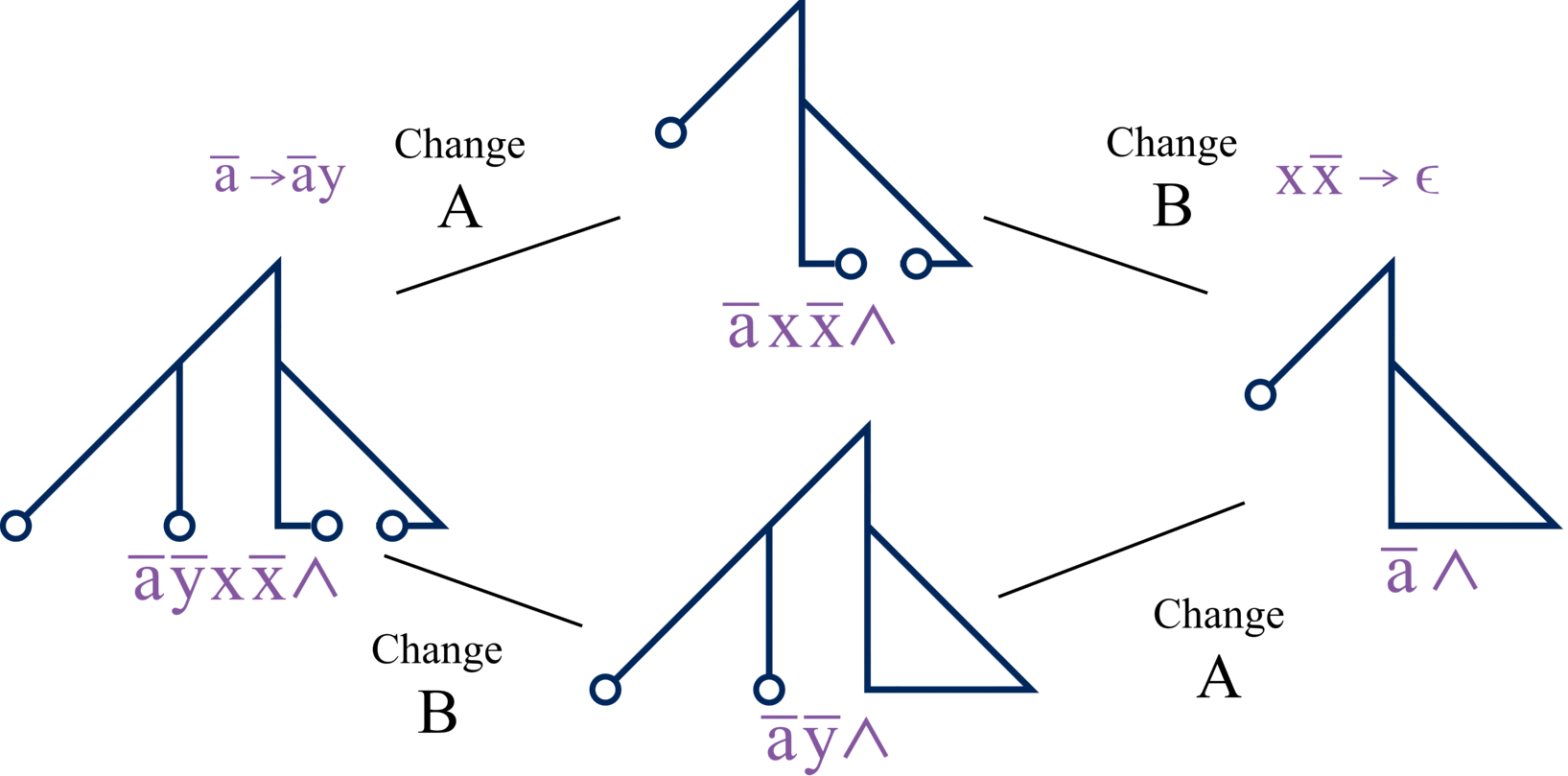}
\end{figure}

The grandchild can be constructed from its grandparent in two different ways. 
We could apply Change A first ($\br{a} \rightarrow \br{ay}$), then Change B
($x\br{x} \rightarrow \epsilon$). Or B first, then A.

When implementing the graph hierarchy, there is a danger of adding different 
copies of a graph to each of its parents. But if we can find all of a graph's 
parents, we can add it as a child of all of them. We find a graph's parents 
by going back to its grandparents and then reversing the order of the two 
operations from grandparent to grandchild (AB to BA). This eliminates the 
possibility of creating duplicate copies of a graph. The above example 
demonstrates how to eliminate duplicates in the case where Change A or B is a
branch gluing operation. But loop gluing can be more complicated.

If a graph has a loop, then the loop can be cut at any of its edges to create 
a different parent graph. The branch gluing operation can be split into two 
parts called half-steps. The first half-step is to add a new primitive. This 
is labeled Change A below. The second step is to glue two half-edges 
together: one from the new primitive and one from the existing graph. This 
second half-step is labeled as B or C below:

\begin{figure}[H]
\centering
\includegraphics[width=4cm]{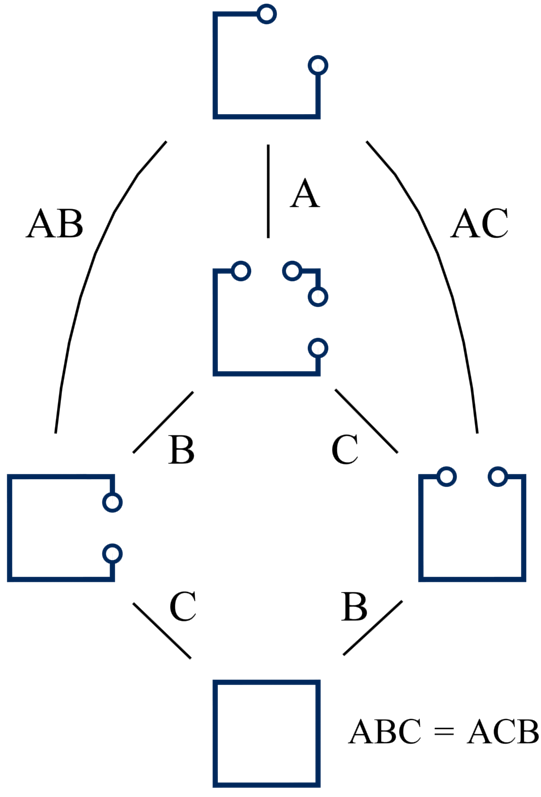}
\end{figure}

Loop gluing is a simpler operation because it only involves the second half-step
of gluing the two half-edges together (Change B or C).

Like before we can eliminate any duplicate graphs by finding all the parents 
of a graph. This is done by reversing the order of the last two half-steps
(BC to CB). As shown above the graph can be constructed by the operation ABC 
or ACB. They are equivalent.

We still must consider the possibility of two consecutive branch gluing
operations. We eliminate duplicates here by reversing the order of the last two
branch gluing operations (each containing two half-steps). As shown below, the
same graph can be constructed through the operations ABAC or ACAB.

\begin{figure}[H]
\centering
\includegraphics[width=5cm]{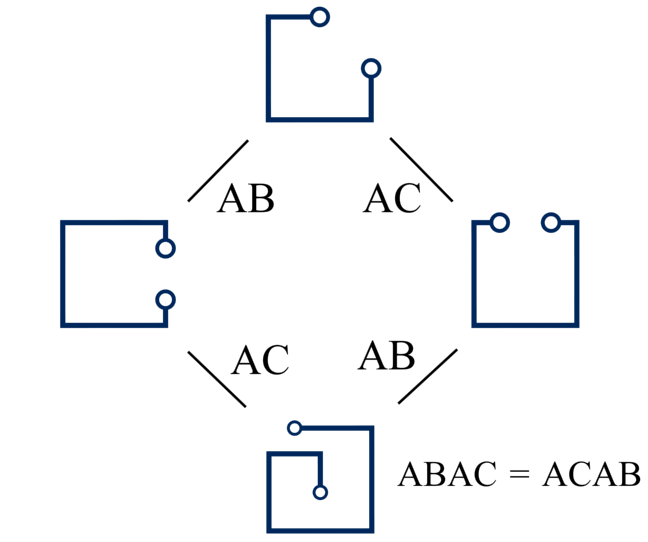}
\end{figure}

\section{Graph Gluing}

\subsection{Loop Gluing Consecutive Edges}

When constructing the graph hierarchy, loop gluing is only allowed between two
consecutive half-edges. This has the effect of removing those two half-edges
from the graph boundary:

\begin{center}
\begin{tabular}{c @{\hskip 1cm} c}
Loop Glue: $a\br{a} \rightarrow \epsilon$ & $\br{a} \bV a\bN \rightarrow \epsilon $ \\
\end{tabular}
\end{center}

It is natural to ask why only consecutive half-edges can be glued together. 
Would allowing non-consecutive loop gluing enable us to construct more locally 
similar shapes that could not be produced otherwise? This section will show 
that the answer is no. Requiring the half-edges to be consecutive does not 
limit the shapes the algorithm produces and simplifies the algorithm.

For example consider the graph below that has two opposite half-edges $\br{b}$
and $b$, but they are not consecutive:

\begin{figure}[H]
\centering
\includegraphics[width=8cm]{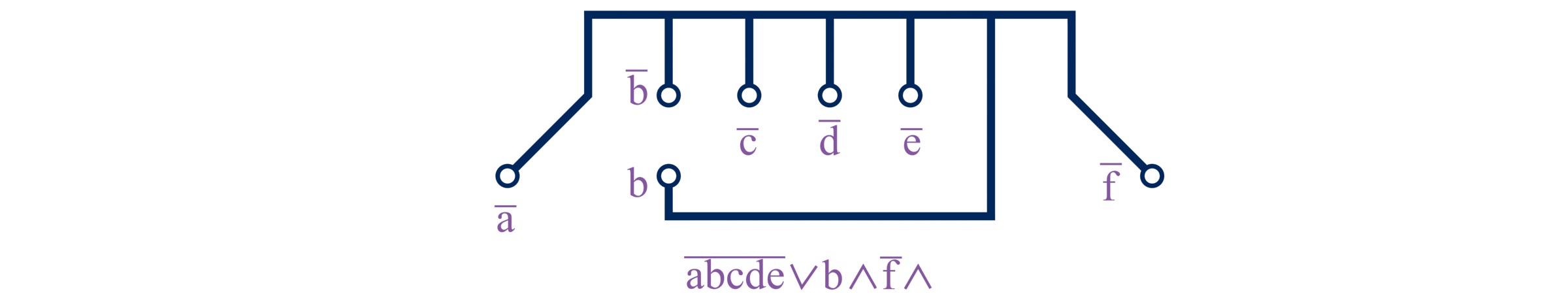}
\end{figure}

Gluing $\br{b}$ and $b$ together results in a graph that can be completed:

\begin{figure}[H]
\centering
\includegraphics[width=8cm]{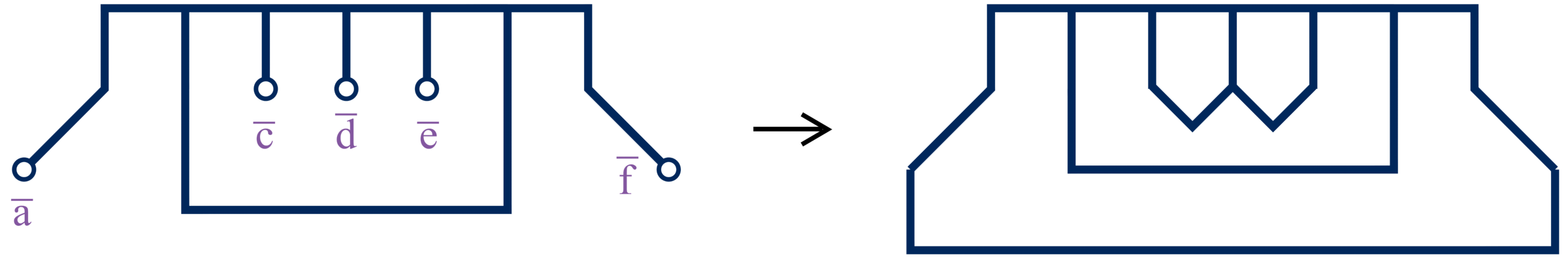}
\end{figure}

But notice that this graph could have been completed by allowing consecutive
half-edges to be loop glued:

\begin{figure}[H]
\centering
\includegraphics[width=8cm]{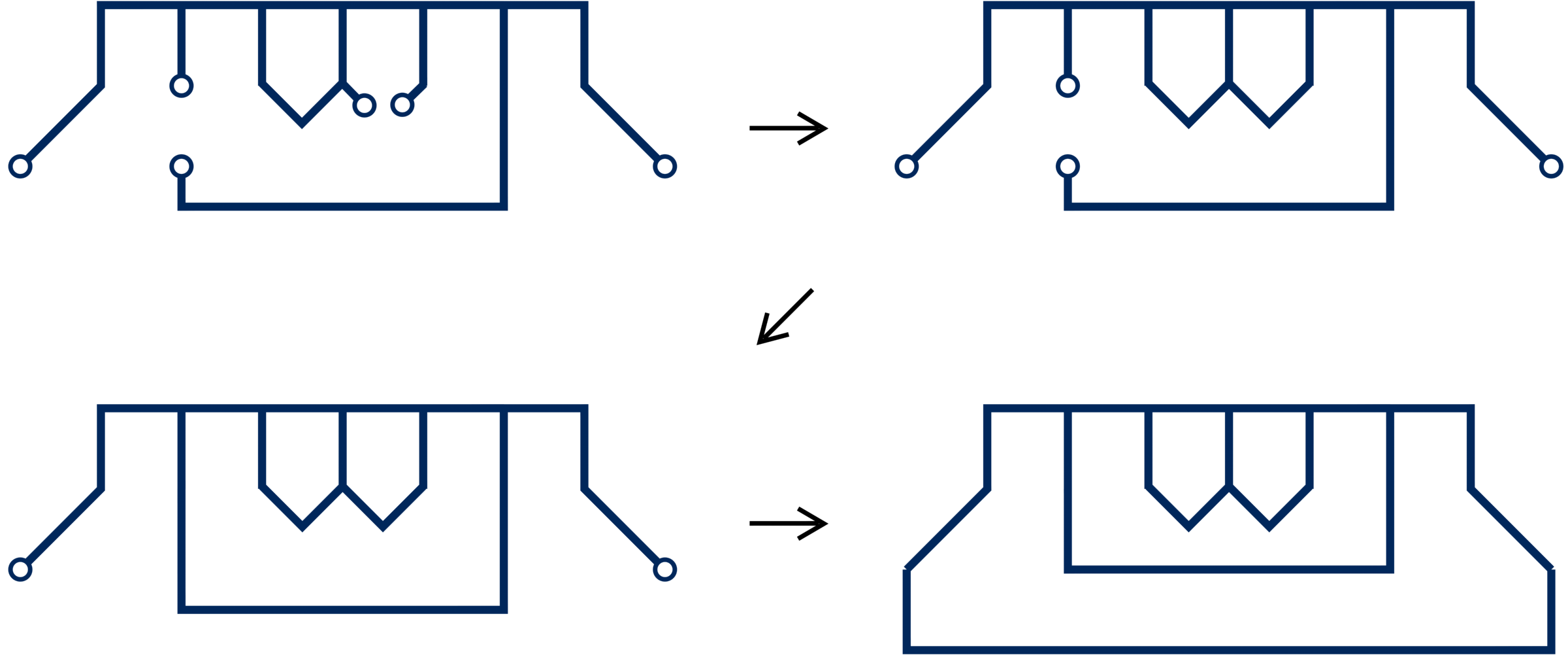}
\end{figure}

If two non-consecutive half-edges are loop glued, this creates a loop that 
divides the remaining half-edges into two categories: those inside the loop
($\br{cde}$) and those outside the loop ($\br{af}$). By only allowing
consecutive half-edges to be glued, we require the interior half-edges
$\br{cde}$ to be glued together first before $\br{b}$ and $b$. Since $\br{cde}$
must eventually be glued together, requiring them to be glued first does not
limit the algorithm. The algorithm can generate the same set of shapes with or
without allowing non-consecutive half-edges to be glued. And the algorithm is
simpler if they are not allowed.

\subsection{Gluing Edge Graphs}
\label{edge gluing}

The simplest graphs in the graph hierarchy are the edges found in Generation 0
of the hierarchy. These graphs consist of a single edge with two half-edges 
pointed in opposite directions. Their graph boundary string always has the 
form $\br {a}a \bN$ for some label $a$. Branch gluing an edge graph to 
another graph has no effect. The general formula for branch gluing $\br{a}B$ 
to $a$ is $a \rightarrow B \bV$. If we glue the edge graph $\br{a}a \bN$ to $ 
a$, the formula becomes $a \rightarrow a \bN \bV$ or $a \rightarrow a$. 
Therefore the gluing has no effect.

Similarly, the general formula for gluing $aB$ to $\br{a}$ is $\br{a} 
\rightarrow \bV B$. If we glue the edge graph $a \bN \br{a}$ to $\br
{a}$, the formula becomes $a \rightarrow \bV \bN  \br{a} = \br{a}$. 
Again gluing an edge graph has no effect.

\subsection{Grouping Vertices}
\label{Grouping Vertices}

Consider this set of four primitives from the diagonal example:

\begin{figure}[H]
\centering
\includegraphics[width=8.0cm]{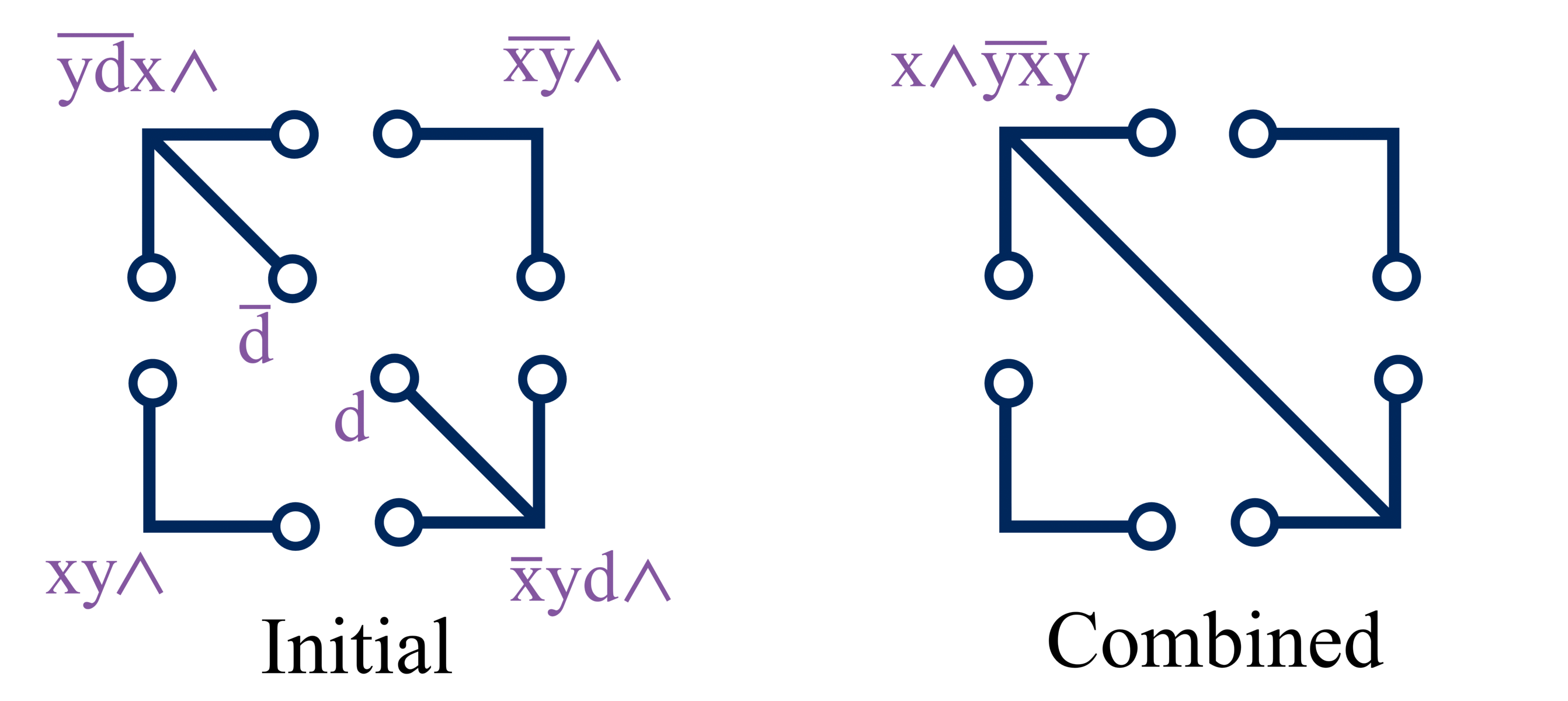}
\end{figure}

The horizontal half-edge $x$ can be glued to one of two primitives $\br{xy} \bN$
or $\br{x}yd \bN$. Each of the half-edges labeled $y$, $\br{x}$, and $\br{y}$ can also be
glued to one of two primitives. But the diagonal half-edges $d$ and $\br{d}$
only have a single choice. $x \bN \br{yd}$ must be glued to $d \bN \br{x}y$. There
are no other options. Since those two primitives must always be glued together,
we can permanently glue them together treating them as if they were a single primitive:
$x \bN \br{y} \bV \bN \br{x} y = x \bN \br{yx} y$.

\begin{table*}[tb]
\begin{center}
\begin{tabular} {lccccccc}
1D & 1D Edges        & divided by & 0D Vertices & & & & \\
2D & 2D Faces        & divided by & 1D Edges    & that intersect at & 0D Vertices & & \\
3D & 3D Volumes      & divided by & 2D Faces    & that intersect at & 1D Edges & that intersect at & 0D Vertices \\
4D & 4D Hypervolumes & divided by & 3D Volumes  & that intersect at & 2D Faces    & that intersect at & 1D Edges \ldots \\
& \vdots &&&&&& \\
\end{tabular}
\caption[]{How the algorithm operates in different dimensions}
\label{dimension}
\end{center}
\end{table*}

We can combine the four initial primitives into three combined primitives as 
shown above. The algorithm derives the same grammar from either set, but it 
is more efficient to combine them.

\subsection{Grouping Edges}
\label{Grouping Edges}

Consider this set of primitives $\mathcal{M}$:

\begin{figure}[H]
\centering
\includegraphics[width=8.0cm]{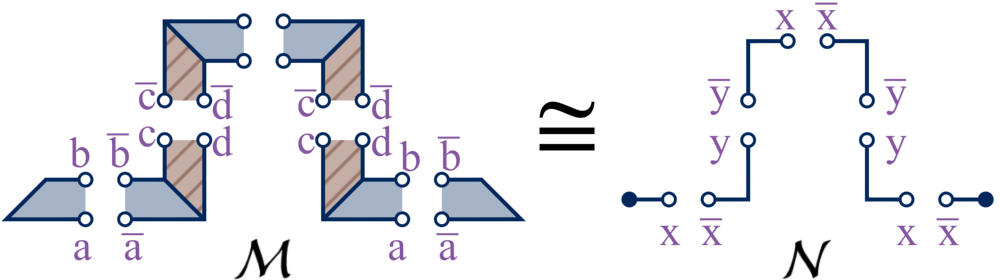}
\end{figure}

Following Section \ref{Grouping Vertices}, we have glued together half-edges that 
must be attached, but we can go further than this. For every primitive in set 
$\mathcal{ M}$, the horizontal and vertical half-edges come in inseparable 
pairs. The horizontal half-edges $(a, b)$ must be glued to an opposite pair
of half-edges $(\br{a}, \br{b})$ on the same primitive. If $a$ is glued to
one primitive and $b$ is glued to another then there is no way to complete the
shape. Splitting the pair in this way mean the path turns more than
$360^{\circ}$ and there is no way to turn it back. (This idea is discussed
more formally in Section \ref{turning upward}. More formally $a$ and $\br{b}$
turn strictly upwards and $\br{a}$ and $b$ turn strictly downward and they
cannot switch directions.)

Each pair of half-edges can only be attached to another pair. The pair of 
half-edges acts as a single unit $(a, b)$, like a single half-edge $x$. The pair
$(c, d)$ acts like a single half-edge $y$ as shown in the figure above. The set
$\mathcal{M}$ is isomorphic to a simpler set, $\mathcal{N}$, in which each pair is 
replaced by a single half-edge. The set $\mathcal{N}$ has fewer half-edges 
and is easier to solve.

\section{Other Dimensions}

This section examines how the algorithm works in different dimensions. In each
dimension, we fill up the space with regions having different labels. In the 2D
case, these are 2D faces with different labels. These n-dimensional regions
have (n-1) dimensional interfaces that divide them and those (n-1) dimensional
interfaces intersect at (n-2) dimensional interfaces and so on (Table
\ref{dimension}).

The 1D version of the algorithm generates a 1D line divided into line 
segments of different colors. The 1D case is easy to solve. The 4D version of 
the algorithm could be used for generating 3D shapes that change over time. Here
are how the graphs and boundaries look in different dimensions:

\begin{figure}[H]
\centering
\includegraphics[width=7cm]{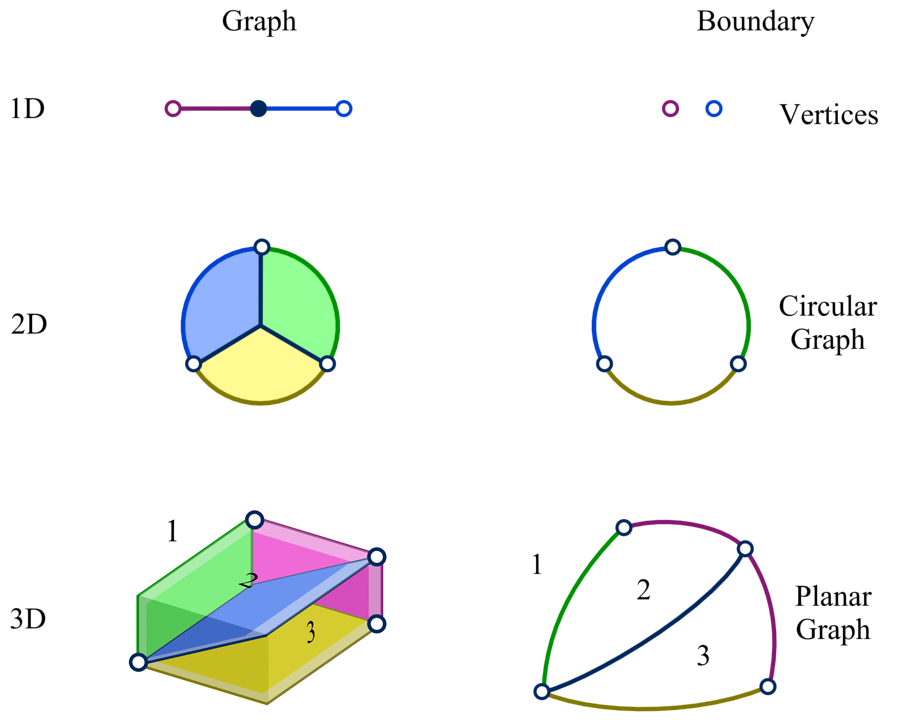}
\end{figure}

We have described the graph boundary as being a 1D string, but it is more
accurate to think of it as a circular graph. The string format is convenient
especially in text form. But it represents a circular path around
the graph. Circularly shifted strings are equivalent:
$y\bN \br{ba} = \bN \br{ba}y = \br{ba}y \bN$. When the
boundary is thought of as a graph, each edge of the graph can be labeled with
the number of positive turns minus negative turns to keep track of that
information.

3D shapes are enclosed by a 2D boundary. Topologically, the 2D boundary is like
a sphere. Any graph that can be drawn on a sphere can also be drawn on a plane.
So the boundary is always a planar graph. Many 3D shapes have boundaries that
are circular graphs which can be described by a 1D string. But this is not
possible if the 3D shape contains any edges that touch three or more faces.

\subsection{Merging and Splicing}

In 2D, there are two simple ways of combining two separate graphs together: (1)
branch gluing where we merge together two half-edges, (2) splicing where we
merge two faces together. Any two faces with the same face label (shown as a
color) can be merged together. The effect on the graph boundary is to splice the
two circular graphs boundaries together which is why it is called a splice.
Splicing is performed by rearranging how the edges connect between the vertices:

\begin{figure}[H]
\centering
\includegraphics[width=5cm]{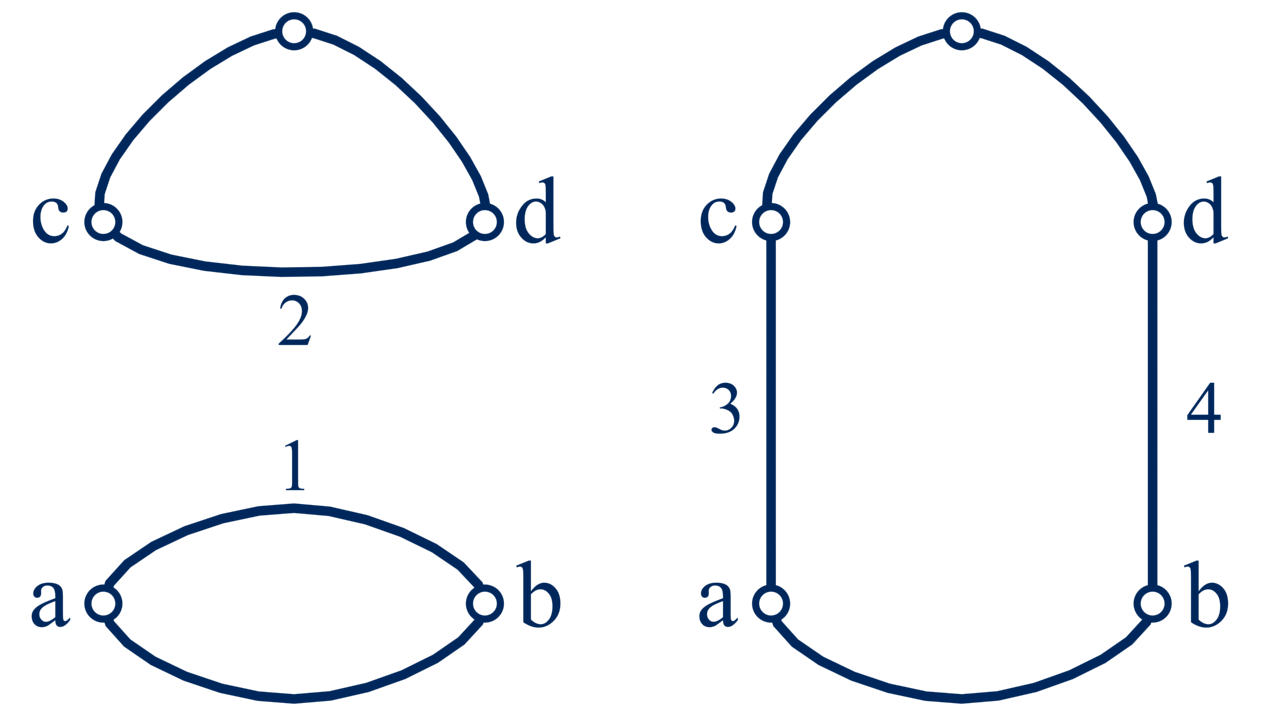}
\end{figure}

Originally, Edge 1 goes from vertex a to b. Edge 2 from c to
d. The splice operation replaces these edges. Edge 3 goes from a
to c. Edge 4 from b to d.

Branch gluing can be thought of as a splice operation followed by a loop gluing
operation. Alternatively, it could be thought of as two splice operations that
create an interior loop that is removed:

\begin{figure}[H]
\centering
\includegraphics[width=8cm]{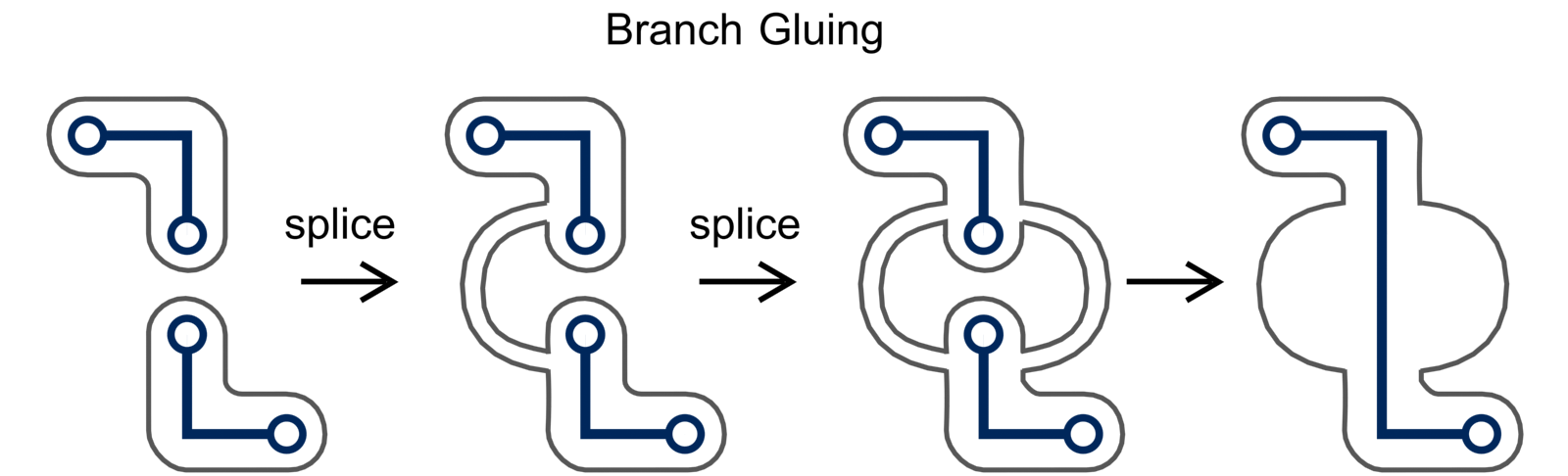}
\end{figure}

In 3D, there are three ways of combining two separate graphs together:
(1) merging edges together, (2) merging faces together, (3) merging volumes
together. This is similar to the 2D case, in that splicing (merging faces) is a
more basic operation than branch gluing (merging edges). Merging volumes is a
more basic operation than merging faces which is more basic than merging edges.
The more basic operation must be performed before the others can be
accomplished. Just as any faces with the same label (shown as a color) can be
merged together, so too can any volumes with the same label be merged together.

\subsection{3D Wireframe}

The 3D algorithm that has been discussed assumes the input and output shape 
consist of 2D polygonal surfaces bounded by edges. But there are other kinds of
3D shapes we could consider. The input and output shape could consists of edges
or line segments with no 2D polygonal surface between them. These would appear
as wireframe objects.

Some of the algorithm could be extended to this use case without difficulty. 
Branch gluing is straightforward, but loop gluing is not. In 2D, loop gluing 
is accomplished by keeping track of positive $\bN$ and negative $\bV$ turns 
to see if a path has turned $360^{\circ}$. But in 3D, the path has more 
degrees of freedom. In 2D, the direction of an edge can be determined by a 
single angle. 2D polar coordinates are defined by a radius and an angle. But 
in 3D, two angles are required to define a direction: a polar angle and an 
azimuthal angle. 3D spherical coordinates are given by a radius and two 
angles. For 3D paths, the simple scheme of using positive and negative 
turns is insufficient. This scheme works fine if the edges are connected to 2D
polygonal surfaces because we can define a coordinate system on each
surface. Allowing us to define positive and negative turns and loop gluing on those 
surfaces. A more sophistical scheme is necessary for 3D wireframe objects 
without surfaces.

The problem of creating 3D wireframe objects that loop is an interesting 
topic for future research. However, this is not a common use case. There are 
many objects that can be well represented as a collection of 3D edges without 
surfaces. A tree or a bush are good examples. However, typically the branches 
of these structures are not allowed to form closed loops, making loop gluing 
unnecessary.

\section{The Boundary Complement}

For every graph boundary string $B$, there exists another boundary string that
could be glued to it to complete the graph. In fact, there are multiple boundary
strings that can complete the graph, one for each half-edge in $B$. We will see
later that this boundary string has the form $\bN B^C D^C \bN D$ for any string
$D$.

It may or may not be possible to construct a graph with the boundary string
$B^C \bN \bN$ using gluing or splicing. If it is possible to construct
such a graph, the graph $B$ can be completed. Otherwise, it is impossible to
complete graph $B$, i.e. nothing can be glued to it to make it complete.

\subsection{Computing the Complement $B^C$}
\label{complement}

We now introduce some new notation. Every boundary string $B$ has a complement
$B^C$. We define the complement to the string that would allow us to simplify
$BB^C$ to the empty string $\epsilon$ using loop gluing operations

\begin{equation}
BB^C \rightarrow \epsilon
\end{equation}

Let us list the various types of half-edge and turns and see what their
complement should be:

\begin{center}
\begin{tabular}{l @{\hskip 1cm} l}
$aa^C=\br{a}a \rightarrow \epsilon $                   & $a^C=\br{a}$         \\
$\br{a}\br{a}^C=\br{a}\bV a \bN \rightarrow \epsilon $ & $\br{a}^C=\bV a \bN$ \\
$\bN \bN^C = \bN \bV \rightarrow \epsilon $            & $ \bN^C=\bV$         \\
$\bV \bV^C = \bV \bN \rightarrow \epsilon $            & $   \bV^C=\bN$
\end{tabular}
\end{center}

When the complement operation is applied to multiple symbols, it reverse the
order: $(AB)^C = B^C A^C$. The complement is applied by reversing the order of
all the symbols and then applying the complement to each individually. For
example, suppose $B = \br{a} \bV b \bN c \bN$:

\begin{equation}
\begin{split}
B^C  & = \bV \br{c} \bV \br{b} \bN \bV a \bN \\
BB^C & = \br{a} \bV b \bN c \br{c} \bV \br{b} \bN \bV a \bN \\
     & \rightarrow \br{a} \bV b \bN \bV \br{b} \bN \bV a \bN \\
     & \rightarrow \br{a} \bV \bN \bV a \bN \\
     & \rightarrow \br{a} \bV a \bN \\
     & \rightarrow \epsilon
\end{split}
\end{equation}

We repeatedly applied the loop gluing formulas to simplify $BB^C$ to the empty
string $\epsilon$. This can be done for any string $B$.

\subsection{Double Complement $(D^C)^C=D$}

Next we show that a boundary string is equal to the complement of its
complement: $(D^C)^C=D$. Let us work through a simple example:

\begin{equation} \label{eq1}
\begin{split}
D    & = a \bN \\
D^C  & = \bV \br{a} \\
(D^C)^C  & = \bV a \bN \bN \\
         & = a \bN \bN \bV \\
         & = a \bN \\
\end{split}
\end{equation}

This relies on the fact that two circularly shifted strings are equivalent 
and that positive and negative turns cancel. Let us work through a more 
complex example to illustrate the principle more generally:

\begin{equation}
\begin{split}
B       & = \br{a} \bV b \bN c \bN \\
B^C     & = \bV \br{c} \bV \br{b} \bN \bV a \bN \\
(B^C)^C & = (\bV \br{a} \bN) \bV (\bV b \bN) \bN (\bV c \bN) \bN \\
        & = \bV \br{a} \bV b \bN c \bN \bN \\
        & = \br{a} \bV b \bN c \bN \bN \bV \\
        & = \br{a} \bV b \bN c \bN \\
        & \rightarrow B \\
\end{split}
\end{equation}

After applying the complement operation twice, each half-edge has a positive 
and negative turn added to it: $(\br{a}^C)^C = \bV \br{a} \bN$ and
$(a^C)^C = \bV a \bN$. But each of the additional turns cancels out with an
additional turn beside the next half-edge or the previous half-edge.

\subsection{Completing the Graph}
\label{completing}

We can complete a graph with the boundary $B$ by gluing it to a graph with the
boundary $ \bN B^C \bN$. We must add two positive turns to $B^C$ because $B^C$ is
not a valid boundary string. If $P(B)$ and $N(B)$ are the number of positive and
negative turns in the string $B$, then $P(B) + N(B) = 1$ for any valid boundary
since the boundary always loops once counter-clockwise around its graph. When we
take the complement, we negate the positive and negative turns, so
$P(B^C) + N(B^C) = -1$. By adding two positive turns, this becomes a valid
boundary string: $P( \bN B^C \bN) + N(\bN B^C \bN) = -1 + 2 = 1$.

A graph with the boundary $\bN B^C \bN$ can be glued to $B$ to complete the 
graph. A complete graph is defined as one with the boundary $\bN$. More 
generally, any graph of the form $\bN B^C D^C \bN D$ can be glued to $B$ to 
complete it where $D$ is any string. More precisely, $\bN B^C D^C \bN D$ is 
branched glued to $B$ and then additional loop gluing is performed. However,
the distinction between branch gluing and loop gluing is not so important since
as described in the original paper, branch gluing is equivalent to a splice 
followed by loop gluing. So this can be described as a splice operation 
followed by loop gluing:

\begin{equation}
\begin{split}
B|\bN B^C D^C \bN D| & \rightarrow BB^C D^C \bN D \\
                     & \rightarrow D^C \bN D \\
                     & = DD^C \bN \\
                     & \rightarrow \bN \\ 
\end{split}
\end{equation}

This is one way to complete a graph with boundary $B$, but there are actually 
several possible ways to do this. The boundary string 
represents a circular graph that can be circularly shifted. And we can splice 
together another graph anywhere along the boundary $B$, not just the end. For
example, below we use the notation $\star_1$ to indicate locations where the
$D^C \bN D \bN$ could be inserted:

\begin{equation}
\begin{split}
B    & = \br{a} \bV b \bN c \bN \\
B^C  & = \bV \br{c} \bV \br{b} \bN \bV a \bN \\
     & = \bV \br{c} \star_1 \bV \br{b} \star_2 a \star_3 \bN \\
\end{split}
\end{equation}

By inserting the $D^C \bN D \bN$ at one of those three locations (one after each of
the three half-edges) and then gluing it to $B$ at that location we can complete
graph $B$. Thus there are three graph boundaries that we could glue to $B$ to
complete it: $\br{c} D^C \bN D \bN \br{b} a$,
$\br{c} \bV \br{b} D^C \bN D \bN \bN a$, and
$\br{c} \bV \br{b} a D^C \bN D \bN$.

\begin{figure}[H]
\centering
\includegraphics[width=8cm]{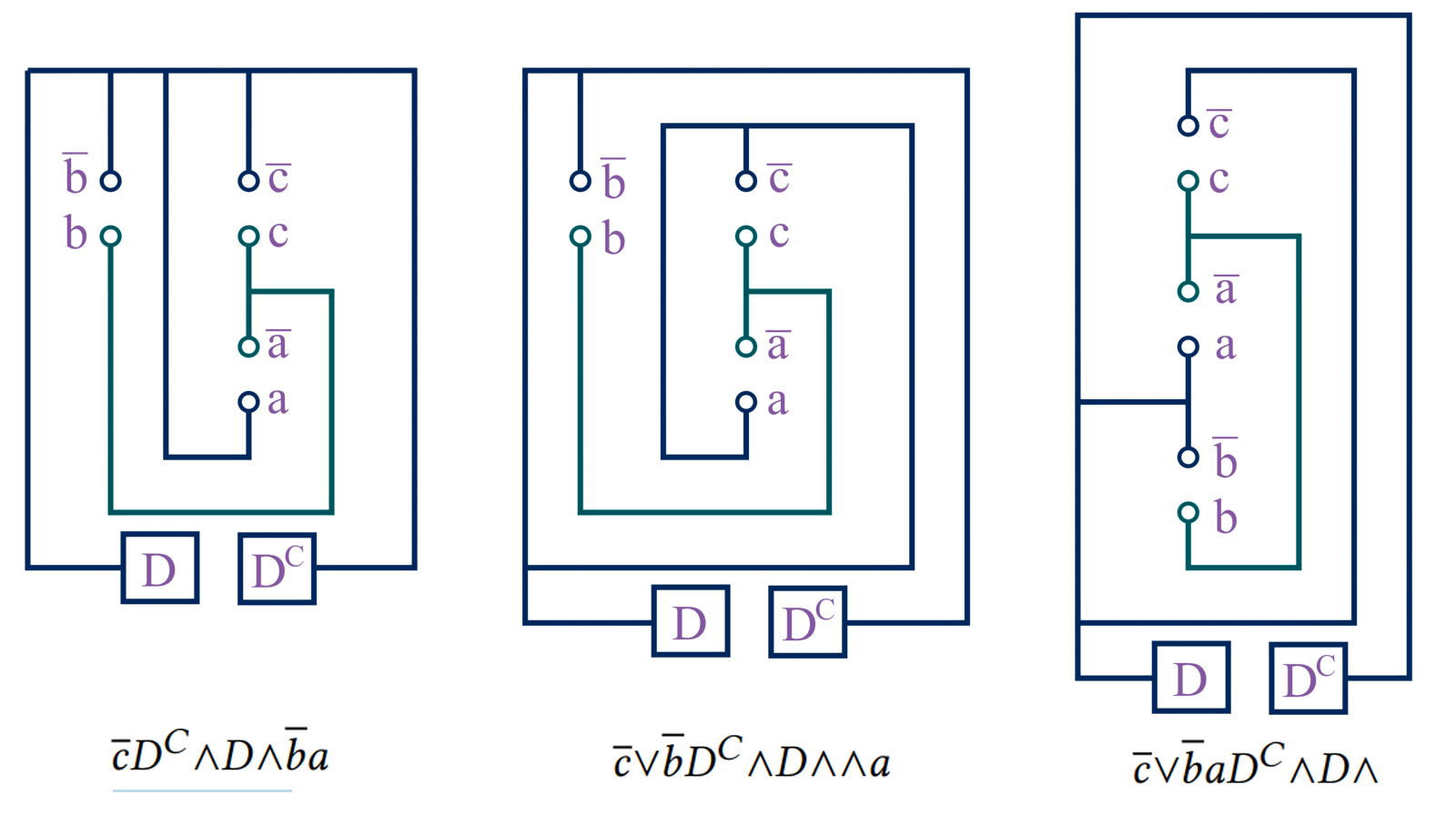}
\end{figure}

And here is another example complementary graph boundaries for a graph with
four half-edges. We have a set of four graph boundaries that can be glued to it:

\begin{figure}[H]
\centering
\includegraphics[width=8cm]{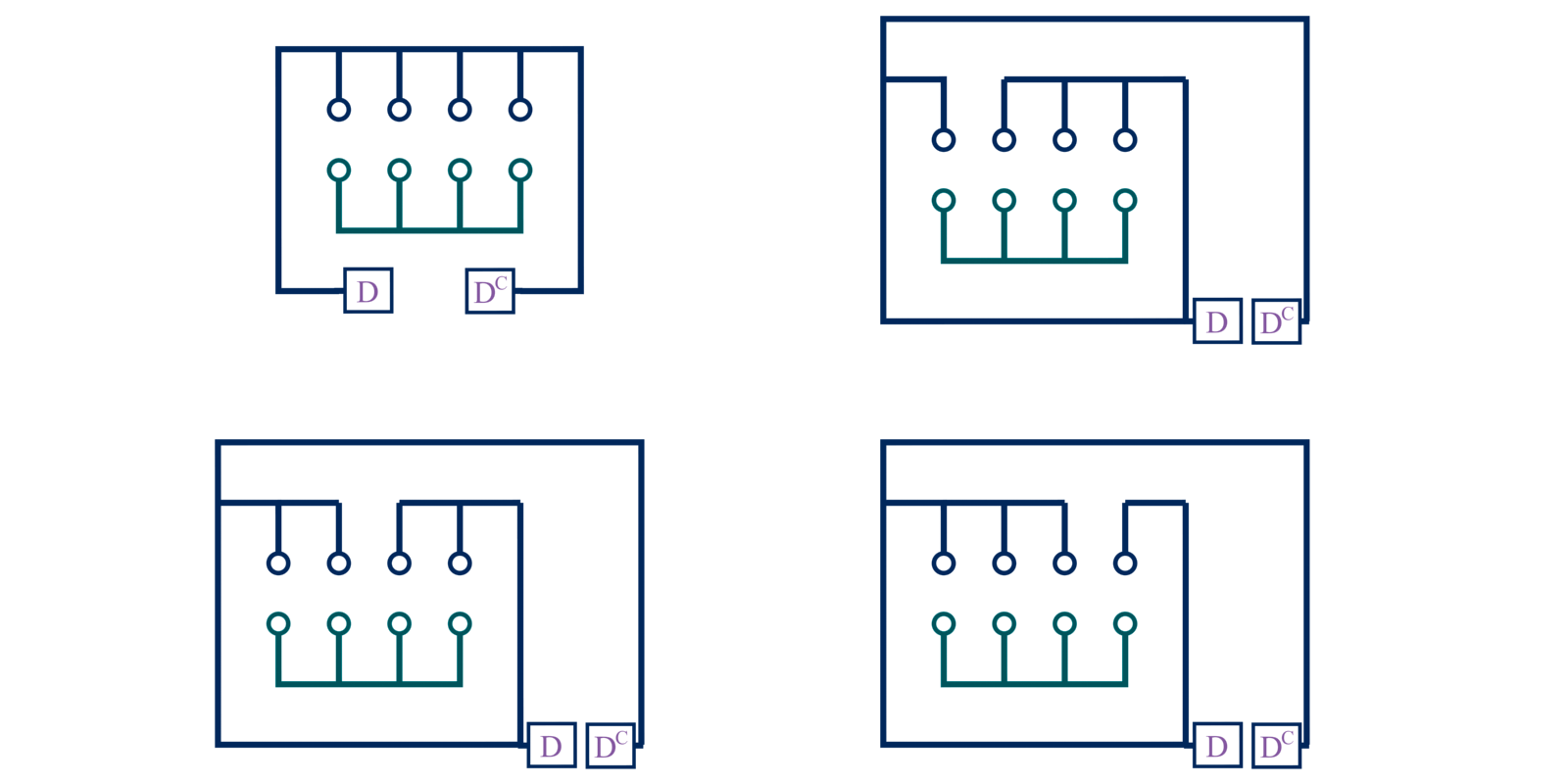}
\end{figure}

\subsection{Completable vs Reducible Graphs}

The problem of determining if a graph is reducible is remarkable similar to the
problem of determining if a graph can be completed. A graph with boundary
$B$ can be reduced if a simpler graph can be constructed with the boundary $B$.
``Constructed'' here means it can be created through some combination of branch
gluing, loop gluing, and splicing. A graph with boundary $B$ can be completed if
a graph can be constructed with the boundary $\bN B^C D^C \bN D$. The same algorithm
outlined as Algorithm 2 in the original paper can be used to determine if such
a graph can be constructed by splicing together the graphs in the hierarchy. It
can be used for determining both reducible and completable graphs.

\subsection{Self-Completable Graphs}

There are some graphs that can be completed without any assistance. They can be
loop glued together by themselves. For example, a graph with the boundary
$B = ab\br{b}c\br{c}\br{a} \bN$ can be completed by three loop gluing
operations. We can find a graph that can complete this graph with the
method described in Section \ref{complement} using $\bN B^C \bN$.
(Here $D = \epsilon$). But that graph has the structure of being a set of
edge graphs $\br{a}{a} \bN$ that are spliced together. Gluing edge graphs to $B$
has no effect (see Section \ref{edge gluing}). So the graph can be completed
using the method of Section \ref{complement}. Even though this is unnecessary
since the graph can be completed without assistance. For example:

\begin{equation}
\begin{split}
B   & = \bN ab\br{b}c\br{c}\br{a} \\
B^C & = \bV ac \bN \br{c} \bV b \bN \br{b} \br{a} \bV \\
\bN B^C \bN & = ac \bN \br{c} \bV b \bN \br{b} \br{a}
\end{split}
\end{equation}

A graph with the boundary $\bN B^C \bN$ can be constructed by splicing together
three edge graphs $\br{a}{a} \bN$, $\br{b}{b} \bN$, and $\br{c}{c} \bN$.

\begin{equation}
\begin{split}
\br{a}a \bN | c \bN \br{c} | & \rightarrow \br{a}a c \bN \br{c} \\
\br{a}a c \bN \br{c} | b \bN \br{b} | & \rightarrow \br{a}ac \bN \br{c} \bV b \bN \br{b} = \bN \bN B^C \\
\end{split}
\end{equation}

\section{The Extended Graph Boundary}

The graph boundary does not fully describe a graph. Multiple graphs can have the
same graph boundary. The extended graph boundary provides more detailed
information about the outer boundary of the graph. Although it does not provide
information about what is inside any loops.

The normal boundary describes just the half-edges, but the extended boundary 
describes the path between them. It is computed similarly by tracing a path 
counter-clockwise around the outside of the graph. In the normal boundary, we 
write down a list of each half-edge and each positive $\bN$ and negative $\bV$
turn found along the path. But for the extended graph boundary, we write down 
a label for every edge we find along the path. We write down what the half-edge
label would be if the edge was cut in half. And this is written down as a 
sequence of rows where each row describes the path between two half-edges in the
normal boundary. For example, the graph below has three half-edges, so the
extended graph boundary contains three rows:

\begin{figure}[H]
\centering
\includegraphics[width=8cm]{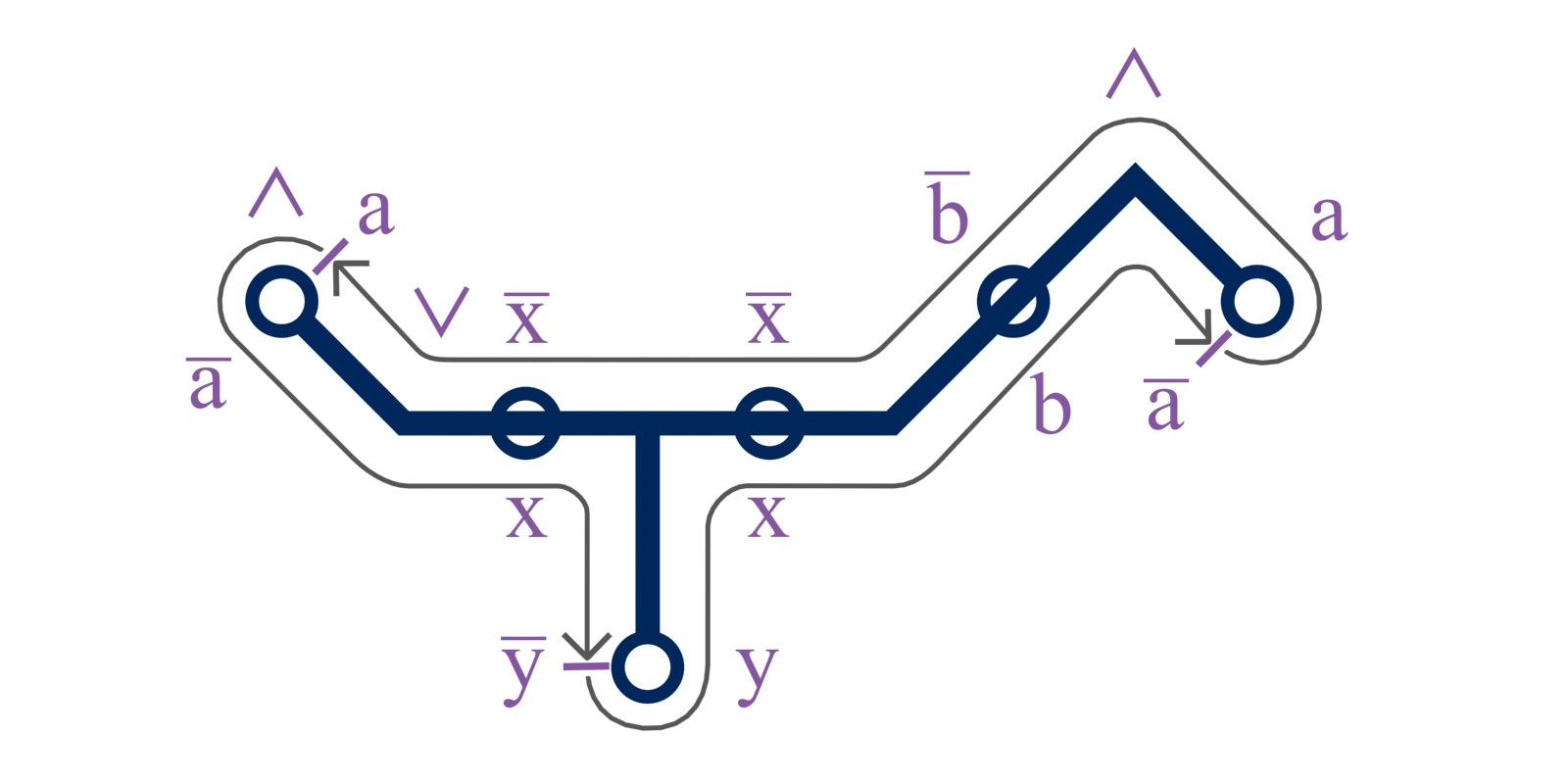}
\end{figure}

\begin{equation}
\label{extended eq}
\begin{split}
a \bN \br{bxx} \bV a \\
\bN \br{a} x \br{y} \\ 
yxb \br{a} \\
\end{split}
\end{equation}

The normal boundary can be derived from the extended boundary by removing every
half-edge except the last one in each row. The normal boundary is 

\begin{equation}
\begin{split}
\bN \bV a \\
\bN \br{y} \\
\br{a} \\
\end{split}
\end{equation}

\noindent which simplifies to $a \bN \br{ya}$. Both boundaries represent 
circular paths. They could be thought of as circular graphs although it is 
convenient to write them as strings. It is arbitrary where the boundary 
starts. It can start at any half-edge.

The last half-edge of each row points in the opposite direction of the first 
half-edge in the next row. In the example above (Eq. label{extended eq}), the
last half-edge in Row 1 is $a$, so the second row begins with $\br{a}$.

The extended boundary can uniquely describe any graph that does not have 
loops. The normal boundary cannot do this. The extended boundary has no 
information about what is inside of any loops. Loops may contain complex 
structures that are not recorded in the extended boundary.

\subsection{Gluing Extended Boundaries}

Now we consider how loop gluing and branch gluing change the extended graph
boundary. Below we see the result of loop gluing:

\begin{figure}[H]
\centering
\includegraphics[width=6cm]{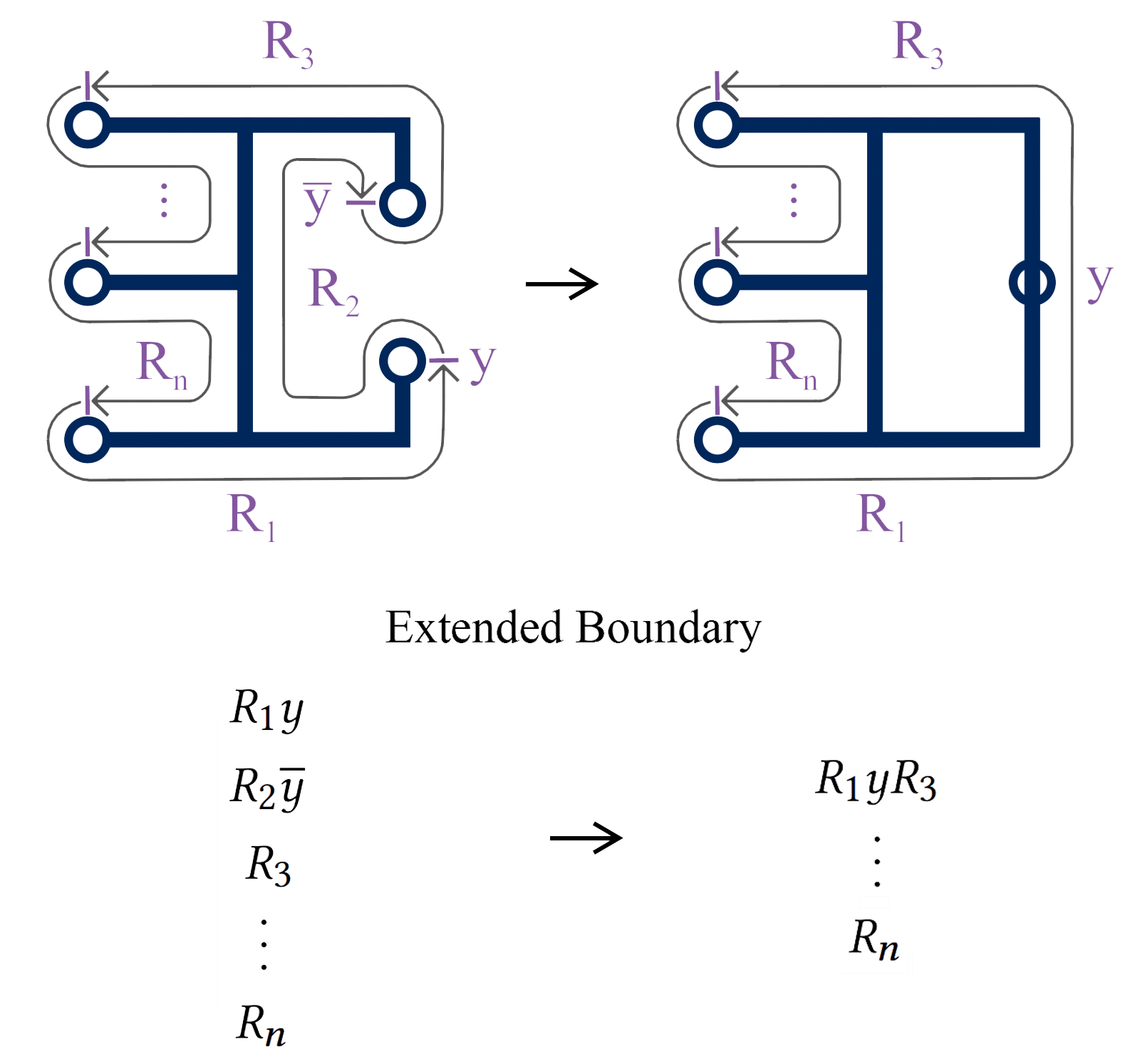}
\end{figure}

The graph has $n$ half-edges and $n$ rows in the extended boundary. The $R_i$
values represent arbitrary strings one on each row. The half-edges $y$ and
$\br{y}$ on rows 1 and 2 are loop glued together. This combines some rows and
decreases the number of rows by 2.

Next consider branch gluing two graphs at the half-edges $a$ and $\br{a}$:

\begin{figure}[H]
\centering
\includegraphics[width=8cm]{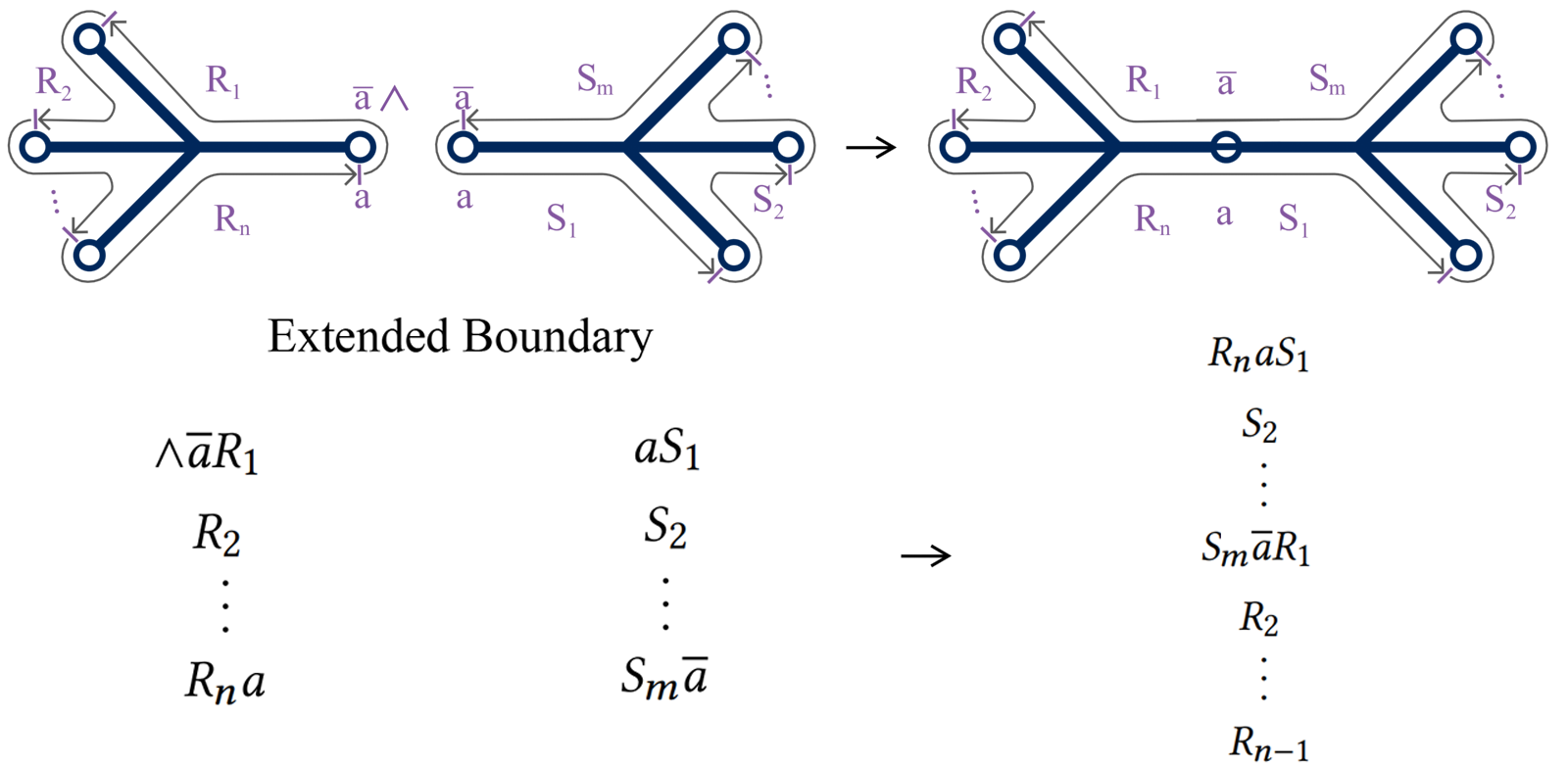}
\end{figure}

The rows for the second graph $(aS_1, S_2, \ldots)$ are inserted into the rows
of the first graph between the last row and the first row.

\subsection{Types of Graphs}

\textit{A complete graph} has no half-edges in its boundary. A complete graph 
always has the boundary $\bN$.

\textit{A stub} is a graph that has one half-edge in its boundary. A stub always has a
boundary of the form $b \bN$ for some half-edge $b$.

\textit{A path graph} is a graph that has two half-edges in its boundary. A path graph
always has a boundary of the form $a \bV^{n-1} b \bN^n$ for two half-edges
labeled $a$ and $b$ and any integer $n$.

\subsection{Extended Boundary of Path Graphs}

If the graph is a path graph and has no loops, the extended boundary has two
rows and the value of the second row can be derived from the first row.

To find the second row from the first, we introduce some new notation. We 
define $\hN$ to be a half-turn $\hN = \bN^{1/2}$. Two half-turns make a full 
turn $\hN\hN = \bN$ and $\hV\hV = \bV$. Then we recursively define the idea
of the transpose of a string $A^T$:

\begin{equation}
\begin{split}
(AB)^T = B^T A^T \\
a^T = \hN \br{a} \hV \\
\br{a}^T = \hN \br{a} \hV \\
\bN^T = \bV = \hV \hV \\
\bV^T = \bN = \hN \hN \\
\end{split}
\end{equation}

We can prove that two transposes always cancel out $(A^T)^T = A$:

\begin{equation}
\begin{split}
((AB)^T)^T & = (B^T A^T)^T \\
           & = (A^T)^T (B^T)^T \\
& \\
(a^T)^T & = (\hN \br{a} \hV)^T \\
        & = \hV^T \br{a}^T \hN^T \\
        & = \hN \br{a}^T \hV \\
        & = \hN \hV a \hN \hV \\
        & = a \\
& \\
(\bN^T)^T & = (\bV)^T = \bV \\
\end{split}
\end{equation}

This proves the two transposes cancel for the $a$ and $\bN$. A similar proof can
be written for $\br{a}$ and $\bV$.

With this new notation, we can derive the second row from the first row. If the
first row has the string $A$, the second row has the string
$A^{-1} = \hN A^T \hN$. For example, if the first row is $\bN \br{b} \bV a$,
then the second row is

\begin{equation}
\begin{split}
(\bN \br{b} \bV a)^{-1} & = \hN (\bN \br{b} \bV a)^{-1} \hN \\
& = \hN \hN \br{a} \hV \hN \hN \hV b \hN \hV \hV \hN \\
& = \bN \br{a} b \\
\end{split}
\end{equation}

\begin{figure}[H]
\centering
\includegraphics[width=5cm]{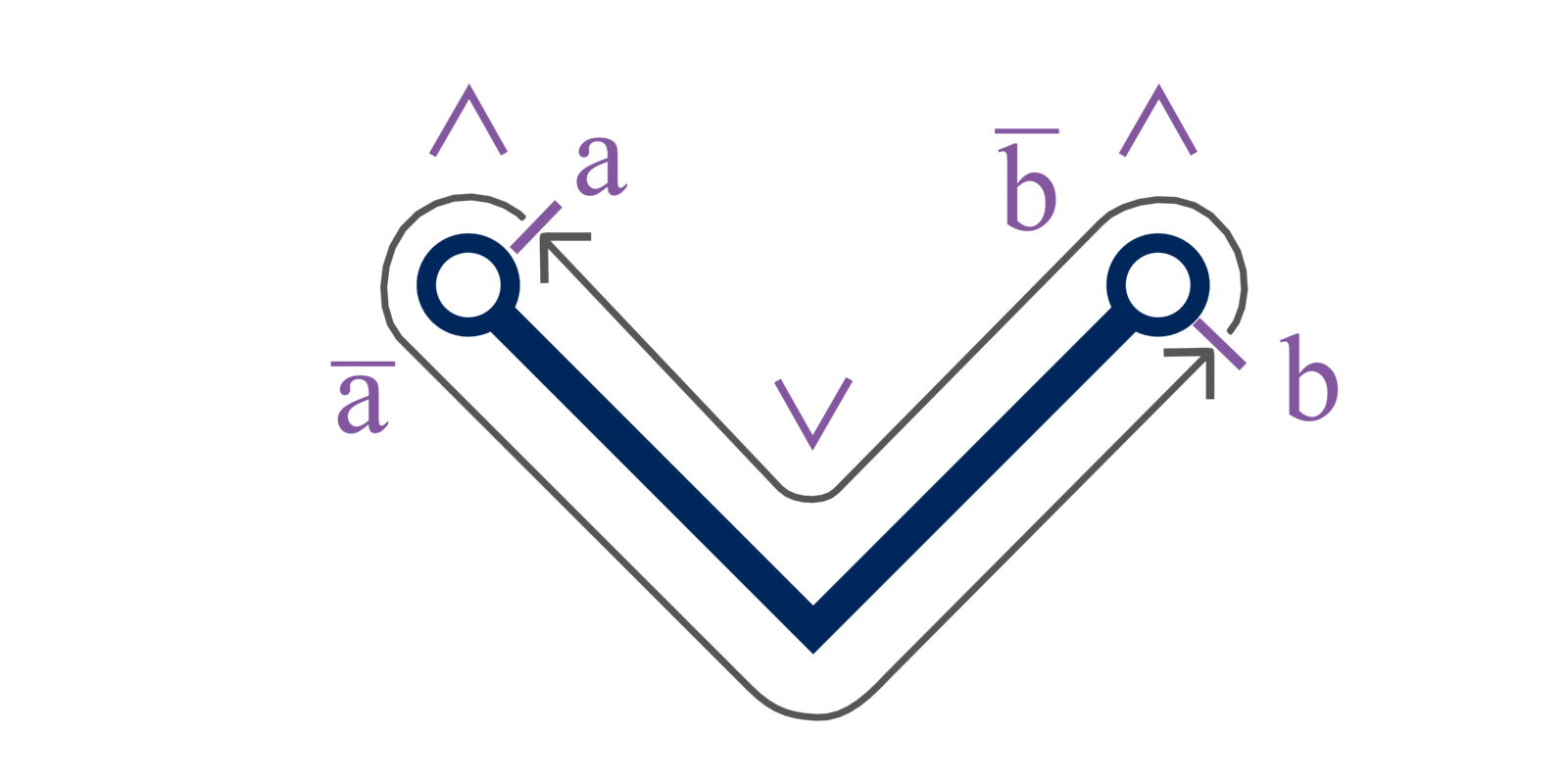}
\end{figure}

We can prove that two inverses cancel $(A^{-1})^{-1} = A$.

\begin{equation}
\begin{split}
(A^{-1})^{-1} & = (\hN A^T \hN)^{-1} \\
              & = \hN (\hN A^T \hN)^{T} \hN \\
              & = \hN \hV (A^T)^T \hV \hN \\
              & = (A^T)^T \\
              & = A \\
\end{split}
\end{equation}

\subsection{Turning Upward and Downward}
\label{turning upward}

We now introduce some notation to describe the extended graph boundary. We use
the term path to describe any part of the path between the half-edges meaning
any row of the extended boundary or any part of a row. The notation

\begin{equation}
a \xrightarrow{n} b
\end{equation}

\noindent describes a path between $a$ and $b$ where $n$ is the number of positive turns
minus negative turns is equal to $n$. For example, the path
$a \bN \bN c d \bV e \bN f b$ can be represented as $a \xrightarrow{2} b$. In
addition to showing the number of turns above the arrow, we will also show them
as subscripts below the half-edges:

\begin{equation}
a_0 \xrightarrow{n} b_n
\end{equation}

We call the subscripts \textit{turning numbers}. The notation $b_n$ means that 
there are $n$ positive turns between $a_0$ and $b_n$. The information in the arrow and 
the turning number are redundant. One value can be determined from the other 
value, but it is helpful to include both. The turning numbers are relative 
values. They are only meaningful by their relationship to other turning numbers.
The turning numbers could all be shifted up or down without changing their meaning. This could have 
been written as $a_{-n} \xrightarrow{n} b_0$. We call the combination $b_n$ a
\textit{turned edge}. It is the combination of a half-edge label and a turning number.

The notation $\left(a_0 \xrightarrow{n} b_n\right)$ this means that there is some way
to construct a path from $a$ to $b$ with $n$ turns. Saying that such a path
``exists'' is different from saying that the ``path exists within a complete
shape''. Some paths can be constructed, but they cannot be formed into a
complete shape.

The arrows are transitive. The end of one path can connect to the beginning of
another:

\begin{equation}
\left(a_0 \xrightarrow{i} b_i \right) \textrm{and} \left(b_i \xrightarrow{j} c_{i+j}\right) \Rightarrow \left(a_0 \xrightarrow{i+j} c_{i+j}\right)
\end{equation}

and we can shift the arrows. For any $n$:

\begin{equation}
\left(a_0 \xrightarrow{i} b_i\right) \Rightarrow \left(a_n \xrightarrow{i} b_{i+n}\right)
\end{equation}

We say that $a$ \textit{turns upward}, if $a_0 \xrightarrow{1} a_1$
exists. Turning upward means turning counterclockwise. The standard convention
for polar coordinates is that angles go up when turning counterclockwise. \newline
We say that $a$ \textit{turns downward}, if $a_0 \xrightarrow{-1} a_{-1}$
exists. Turning downward means turning clockwise. \newline
We say that $a$ \textit{turns strictly upward}, if $a_0 \xrightarrow{1} a_1$ exists and
$a_0 \xrightarrow{-1} a_{-1}$ does not. \newline
We say that $a$ \textit{turns strictly downward}, if $a_0 \xrightarrow{-1} a_{-1}$ exists and
$a_0 \xrightarrow{1} b_{1}$ does not. \newline
We say that $a$ \textit{turns upward and downward}, if $a_0 \xrightarrow{1} b_{1}$
and $a_0 \xrightarrow{-1} a_{-1}$ exists. \newline

If $a$ is part of a complete shape it must turn upward or turn downward or both. The
only way to form a complete shape is to form a path with a $\pm 360^{\circ}$.
In a complete graph, every path forms a loop either around the outside of the
complete graph in as an interior loop. This means that either $a_0 \xrightarrow{1} a_1$ or
$a_0 \xrightarrow{-1} a_{-1}$. (If we started with a complete shape and cut it
into primitives, then every half-edge can be formed into a complete shape.
However, this is not necessarily be true if we are gluing together an arbitrary set of
primitives.)

If $a$ turns upward, then it can continuing turning upward indefinitely. For any positive $n$:

\begin{equation}
\left(a_0 \xrightarrow{1} a_1\right) \Rightarrow \left(a_0 \xrightarrow{n} a_{n}\right)
\end{equation}

If $a$ turns strictly upward, then it can never turn downward any amount. For
any positive $n$, $a_0 \xrightarrow{-n} a_{-n}$ does not exist. If it did exist then:

\begin{equation}
\label{nonnegative}
a_0 \xrightarrow{-n} a_-n \xrightarrow{n-1} a_{-1}
\end{equation}

\noindent contradicting the premise that $a$ turns \textit{strictly} upward.

If half-edges $a$ and $b$ are part of a closed loop this implies that either
$a_0 \xrightarrow{i} b_i \xrightarrow{ 1-i} a_1$ or
$a_0 \xrightarrow{i} b_i \xrightarrow{-1-i} a_{-1}$ for some $i$. If $a$ and $b$
are part of a loop then $a$ turns upward if and only if $b$ turns upward and
$a$ turns downward if and only if $b$ turns downward. Let us assume that
$a_0 \xrightarrow{i} b_i \xrightarrow{ 1-i} a_1$. The following argument would
be similar if we instead assumed that
$a_0 \xrightarrow{i} b_i \xrightarrow{-1-i} a_{-1}$.
The fact that $a$ turns upward implies that $b$ turns upward:

\begin{equation}
\begin{split}
a_0 \xrightarrow{i} b_i \xrightarrow{1-i} a_1 \xrightarrow{i} b_{i+1} \\
b_i \xrightarrow{1} b_{i+1}
\end{split}
\end{equation}

If $a$ turns downward that would imply that $b$ turns downward as well:

\begin{equation}
\begin{split}
b_i \xrightarrow{1-i} a_1 \xrightarrow{-1} a_0 \xrightarrow{-1} a_{-1} \xrightarrow{i} b_{i-1} \\
b_0 \xrightarrow{-1} b_{-1}
\end{split}
\end{equation}

If $b$ turns downward that would imply that $a$ turns downward as well:

\begin{equation}
\begin{split}
a_0 \xrightarrow{i} b_i \xrightarrow{-1} b_{i-1} \xrightarrow{-1} b_{i-2} \xrightarrow{1-i} a_{-1} \\
a_0 \xrightarrow{-1} a_{-1}
\end{split}
\end{equation}

If $a$ and $b$ are part of a closed loop, they must turn the same way.
This argument can be applied to every half-edge in the loop. A closed loop
has three possible options either all the half-edges turn strictly upward, all
turn strictly downward, or all turn upward and downward.

\subsection{The Directed Graph}

It is not difficult to calculate if each half-edge turns upward and / or
downward. We can construct a directed graph where each node is a half-edge. A
directed edge leads from one node to another node if one half-edge can lead to
the next half-edge in the extended boundary. For example,
from the example shape and primitives below we can compute a directed graph:

\begin{figure}[H]
\centering
\includegraphics[width=8cm]{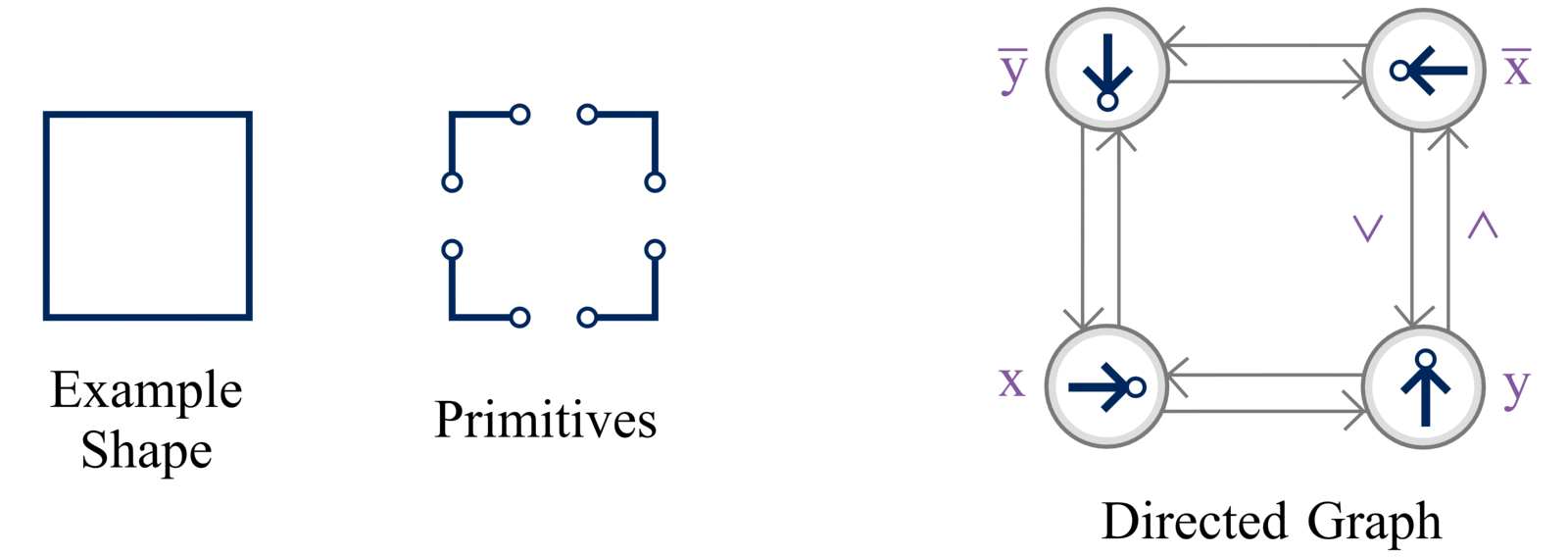}
\caption{A Directed Graph Derived From an Example Shape}
\label{directSingle}
\end{figure}

Notice that the edges are marked with positive turns $\bN$ if traversing them
increments the turning number and negative turns $\bN$ if it decrements the
turning number. A half-edge $x$ turns upwards if a cycle exists that
starts and ends with $x$ and has one positive turn. It turns downward if
such a cycle exists with one negative turn. In Figure \ref{directSingle}, all
the half-edges turn upward and downward.

Cycles with one positive or one negative turns are the only paths allowed in 
a complete graph. The positive cycles correspond to closed loops around the
exterior of the graph. The negative cycles correspond to closed loops in the 
interior of the graph.

Knowing if a half-edge turns upward or downward is very useful because it may 
tell us that a graph cannot be completed. If a graph cannot be completed, it 
and all of its descendants can be removed from the hierarchy. Removing such 
graphs is often necessary for the algorithm to finish. If a half-edge $x$ 
turns strictly upward  and a graph's extended boundary contains any paths
$x_0 \xrightarrow{2} x_{2}$ then the graph cannot be completed. In other words, 
if a half-edge turns strictly counterclockwise, then it cannot be finished if it
makes two turns because there is no way to turn it back.

\subsection{Directed Graph with Pairs of Half-Edges}

We can do something more sophisticated if instead of considering each half-edge
individually, we consider them in pairs. The half-edge $\br{x}$ can be followed
by $\br{xy}$ or $\br{x}y$. We will treat these as being two different options.
We will create a new directed graph where the nodes are pairs of half-edges that
can be connected together:

\begin{figure}[H]
\centering
\includegraphics[width=8cm]{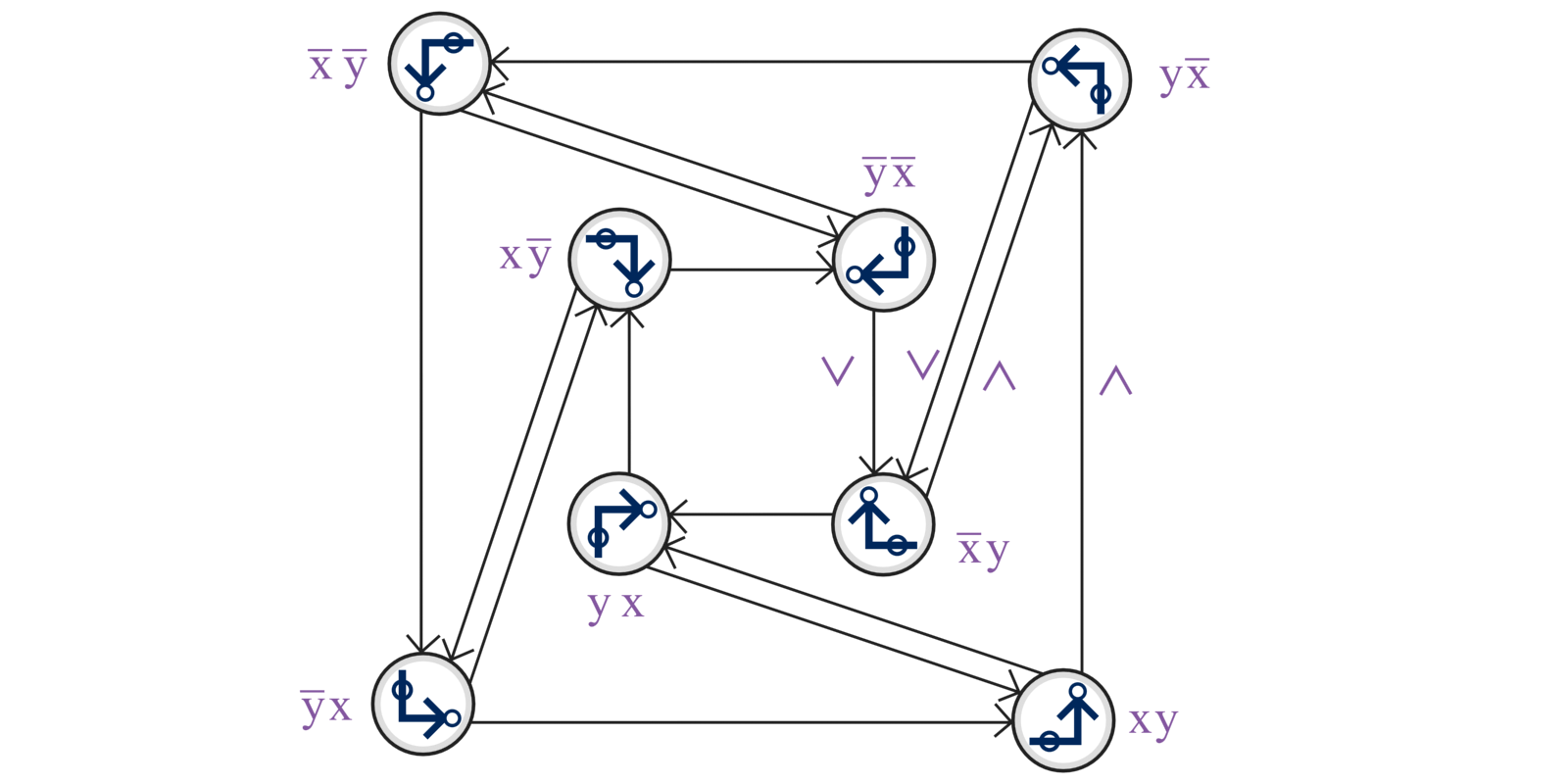}
\caption{A Directed Graph with a Half-Edge Pair at Each node}
\label{directPair}
\end{figure}

The directed edges of the graph tell us which nodes lead to other nodes. For example, the node
$y\br{x}$ has two edges coming from it. One edge points to $\br{xy}$ and one
points to $\br{x}y$. Those are the possible pairs that can follow $y\br{x}$ in a
path.

Again we can construct valid paths by traversing this
directed graph. The same paths can be constructed whether or not we
traverse the directed graph in Figure \ref{directSingle} or the graph in Figure
\ref{directPair}. Valid cycles start and end at the same node with one positive
or one negative turn.

What makes this interesting is when we consider the rules of a grammar. We can
find the following grammar rules using my algorithm:

\begin{figure}[H]
\centering
\includegraphics[width=8cm]{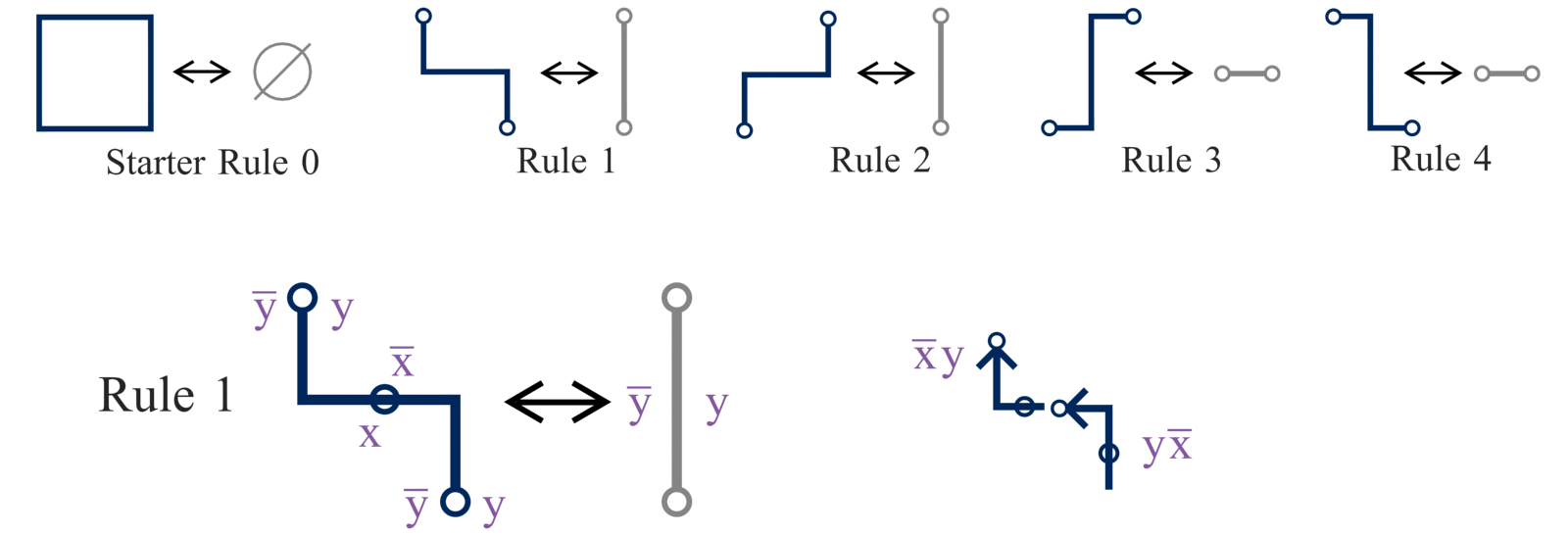}
\end{figure}

If we apply Rule 1 to a graph, we can take any path of the form $y\br{x}y$ and
simplify it to a straight line $y$. The rules consist of graphs having two half-edges and
two paths between them. Each rule can reduce two paths. Rule 1 can reduce
$y\br{x}y$ to $y$ and $\br{y}x\br{y}$ to $\br{y}$. By applying Rule 1, we can
simplify every graph with these paths ($y\br{x}y$ and $\br{y}x\br{y}$). We can
label the directed edges based on which rule eliminates them:

\begin{figure}[H]
\centering
\includegraphics[width=8.5cm]{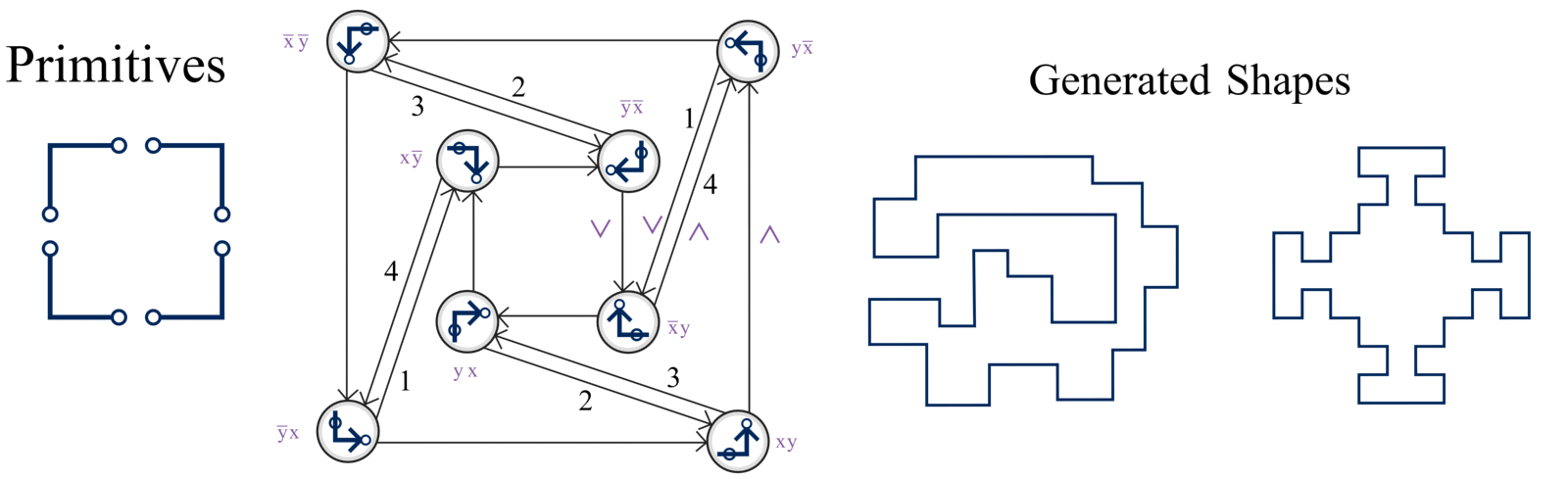}
\end{figure}

Imagine we take the set of all locally similar graphs and reduce them using Rule
1. What is the set of graphs after applying Rule 1? In other words, what graphs
are not reducible by Rule 1? This set of graphs consist of those that can be
constructed by finding cycles in this new directed graph:

\begin{figure}[H]
\centering
\includegraphics[width=8.5cm]{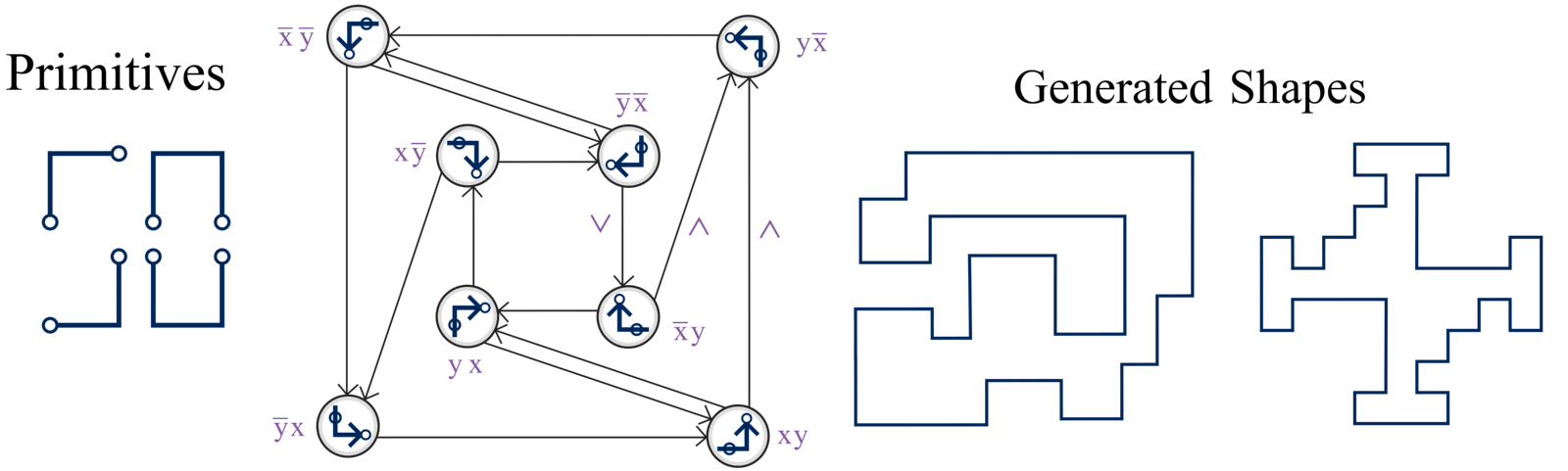}
\end{figure}

This new directed graph is missing the two edges that were eliminated by Rule 1.
We could also describe this set of graph by saying it is the set that can be
constructed using the set of primitives shown in each of the diagrams.

Now let's consider the set of all graphs after applying Rules 1 and 2:

\begin{figure}[H]
\centering
\includegraphics[width=8.5cm]{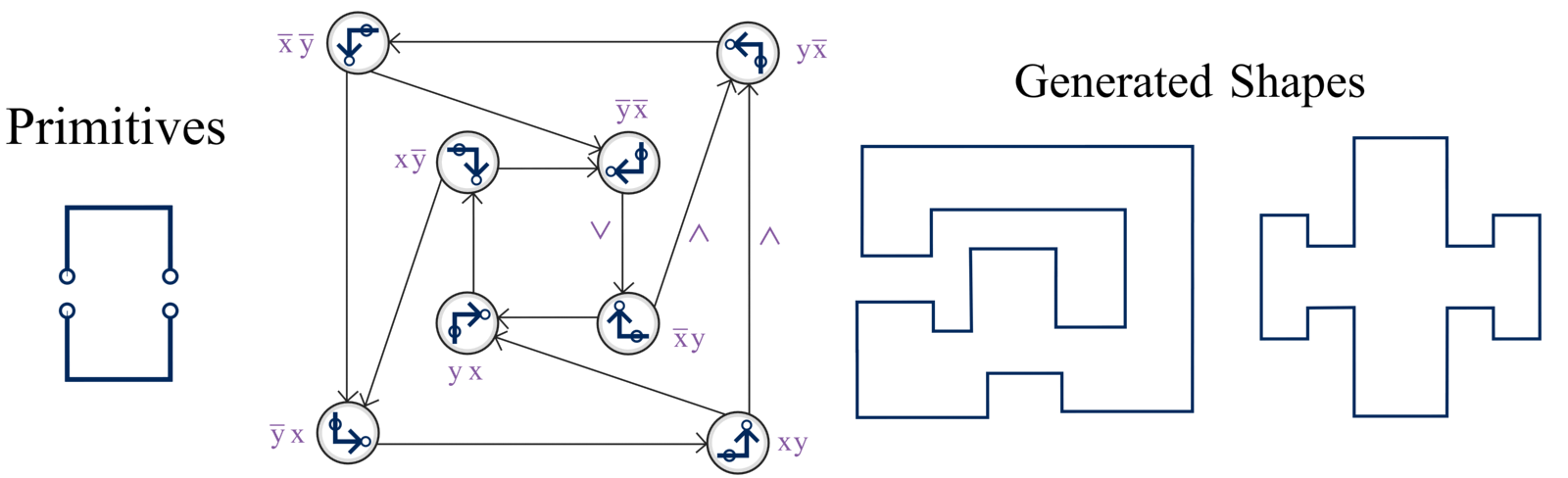}
\end{figure}

Next consider the set of all graphs after applying Rules 1 - 3:

\begin{figure}[H]
\centering
\includegraphics[width=8.5cm]{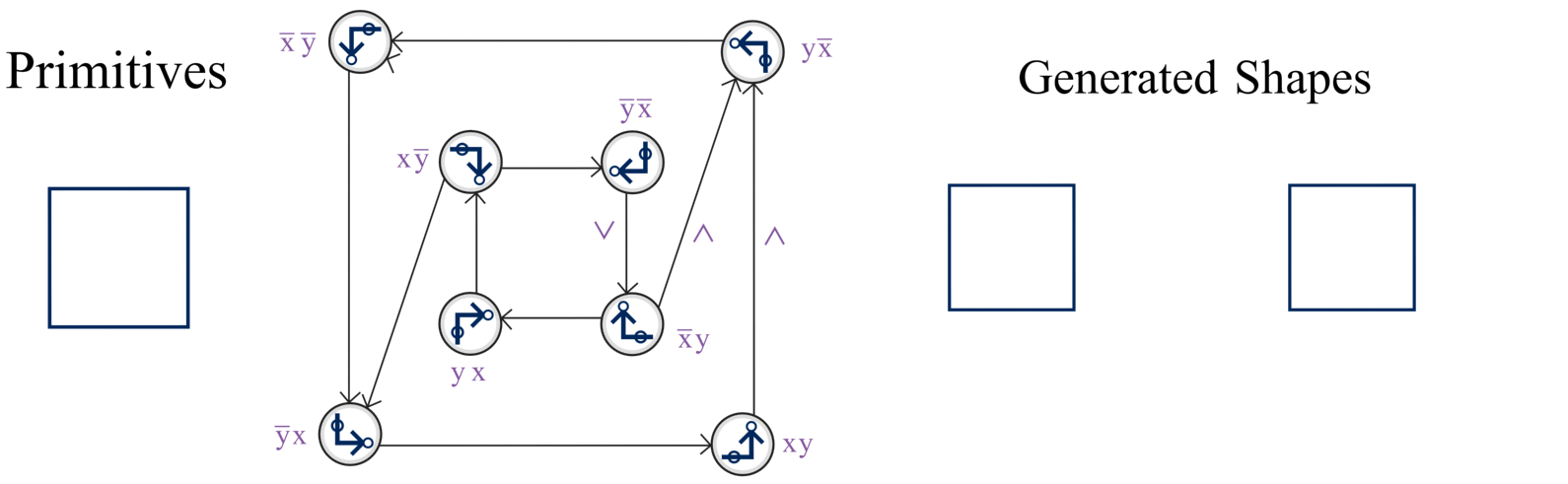}
\end{figure}

We have reduced the set of possible graphs to just one graph which is a simple rectangle.
The directed graph has only two cycles. It has an outer cycle where the path
turns counterclockwise (strictly upward). And it has an inner cycle
which turns clockwise (strictly downward). These two paths correspond to the
closed paths that are inside and outside of a simple rectangle. It is possible to travel
from the inner loop to the outer loop in the directed graph (there is an edge
from $x\br{y}$ to $\br{y}x$). But it is impossible to
go back from the outer loop to the inner loop. So only two cycles exist.

There is one remaining graph that has not been reduced, the simple rectangle. It can be reduced using
the starter Rule 0. Every complete graph can be reduced by Rules 0 - 3. Rule 4
is actually not necessary. This type of directed graph with pair of half-edges
allows us to analyze the set of complete graphs that exist after applying
different rules.

\section{Solution for Path Graphs and Stubs}

\subsection{The Constructable Set $C_{\mathcal{P}}$}

Let $P$ be a finite set of graphs. Let $C_{\mathcal{P}}$ be the set of 
complete graphs that is constructable from $\mathcal{P}$. A graph $G_A$ is
constructable from $\mathcal{P}$ if there is any way to glue the graphs in
$\mathcal{P}$ together to produce $G_A$. Let $C^{\star}_{\mathcal{P}}$ be the
set of graphs in $C_{\mathcal{P}}$ that are complete.

There is another way we can construct graphs. They can be constructed using a
graph grammar. Let $\mathcal{G}$ be a graph grammar. Let
$C^{\star}_{\mathcal{G}}$ be the set of graphs that can be constructed by using
the rules of the grammar. Note that all graphs that can be constructed by a
grammar are complete.

A grammar is \textit{finite} if the number rules it contains is finite.

\subsection{Solution Proposition}

Proposition 1: If every graph in $\mathcal{P}$ has 2 is a path graph of a stub,
then there exists a finite grammar $\mathcal{G}$ such that
$C^{\star}_{\mathcal{P}} = C^{\star}_{\mathcal{G}}$.

Or more generally, let $\mathcal{P}'$ be a finite set of graphs. (Some of the
graphs may not be path graphs or stubs). If there exists a set of rules to
reduce the set $\mathcal{P}'$ to a set $\mathcal{P}$ that only contains path
graphs or stubs, there exists a finite grammar $\mathcal{G}$ such that
$C^{\star}_{\mathcal{P'}} = C^{\star}_{\mathcal{G}}$.

This proposition is important because it says that we can find a perfect
solution in the case of path graphs and stubs. We can find a graph grammar that
generates every locally similar shape. The difficult hard is in ensuring that
the grammar is finite. An infinite grammar can easily solve the problem.

\subsection{The Set $C_{\mathcal{P}}$ for Path Graphs}

Every graph in $C_{\mathcal{P}}$ has 2 half-edges or less. To prove this,
suppose we branch glue two graphs together $G_1$ and $G_2$ and the number of
half-edges they have $h_1$ and $h_2$ respectively. Then the number of
half-edges after gluing $G_1$ and $G_2$ together is $h_1 + h_2 - 2$. Since
$h_1 \leq 2$ and $h_2 \leq 2$, then $h_1 + h_2 - 2 \leq 2$. The number of
half-edges cannot be increased by branch gluing nor by loop gluing. The number
of half-edges cannot be increased beyond 2.

The result of gluing any two path graphs together is another path graph since
$h_1 = h_2 = 2$ implies that $h_1 + h_2 - 2 = 2$. Path graphs and stubs can only
be glued into complete graphs in one of two ways. One option is to create a
simple path with dead-ends at each of its ends:

\begin{figure}[H]
\centering
\includegraphics[width=8cm]{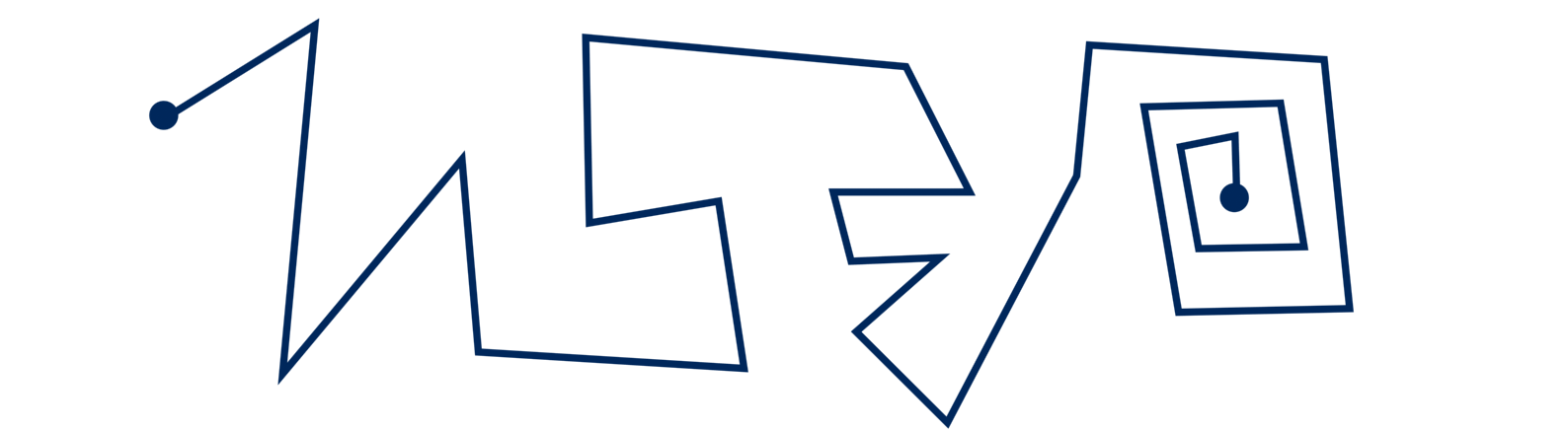}
\caption{A simple path with dead-ends.}
\label{stub path}
\end{figure}

Or a simple closed loop:

\begin{figure}[H]
\centering
\includegraphics[width=8cm]{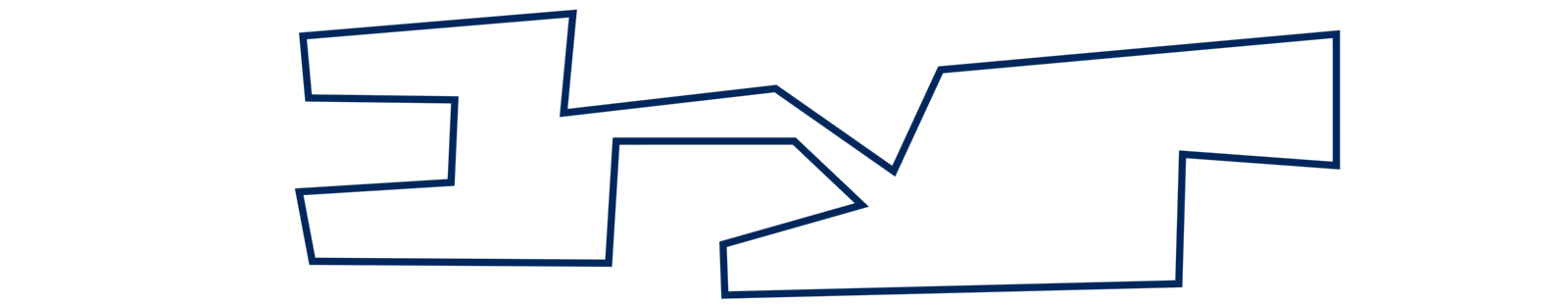}
\caption{A closed loop.}
\label{loop path}
\end{figure}

\subsection{Stubs}

If two stubs $b \bN$ and $\br{b} \bN$ are in $C_{\mathcal{P}}$ and a path exists
from a half-edge labeled $a$ to $b$ or $\br{b}$, then the stub $a \bN$ is also
in $C_{\mathcal{P}}$. If a path exists from $a$ to $b$ that means the graph
$a \bN^n b \bV^{n-1}$ exists and it can be glued to $\br{b} \bN$ to construct
the stub $a \bN$.

If $b \bN$ is in $C_{\mathcal{P}}$, then the stub $\br{b} \bN$ must also be in
$C_{\mathcal{P}}$ exist for $b$ to be part of a complete shape. A stub can only
be completed by gluing it to another stub. To prove this, recall the formula for
completing a graph from Section \ref{completing}. A graph $B$ can only be
completed by gluing it to a graph with the boundary $\bN B^C D^C \bN D$.
We know that only graphs with 2 or less half-edges are in
$C_{\mathcal{P}}$. So the string $D$ must be empty $D = \epsilon$. Otherwise,
$\bN B^C D^C \bN D$ would contain more than 2 half-edges. So the stub $b \bN$
can only be completed by gluing it to the stub $\br{b} \bN$.

If the stub $\br{b} \bN$ is not in $C_{\mathcal{P}}$, then there is no way to
complete the stub $b \bN$. The stub $\br{b} \bN$ must be in $C_{\mathcal{P}}$
if the graphs $\mathcal{P}$ were found by cutting a complete graph into
primitives. But in the general case where $\mathcal{P}$ may contain any graphs -
not just primitives - $\br{b} \bN$ might not be in $C_{\mathcal{P}}$.
If $\br{b} \bN$ is not in $C_{\mathcal{P}}$, then any graph that contains the
half-edges $b$ or $\br{b}$ cannot be completed. Graphs that cannot be completed
need not concern us as Proposition 1 is a statement about the complete graphs of
$C^{\star}_{\mathcal{P}}$.

If $C_{\mathcal{P}}$ contains a path from $a$ to $b$ and the stub $\br{b} \bN$,
then we can glue them together to create the stub $a \bN$. And the stub
$\br{a} \bN$ must also be in $C_{\mathcal{P}}$ for $a$ to be part of a complete
shape.

\begin{figure}[H]
\centering
\includegraphics[width=8cm]{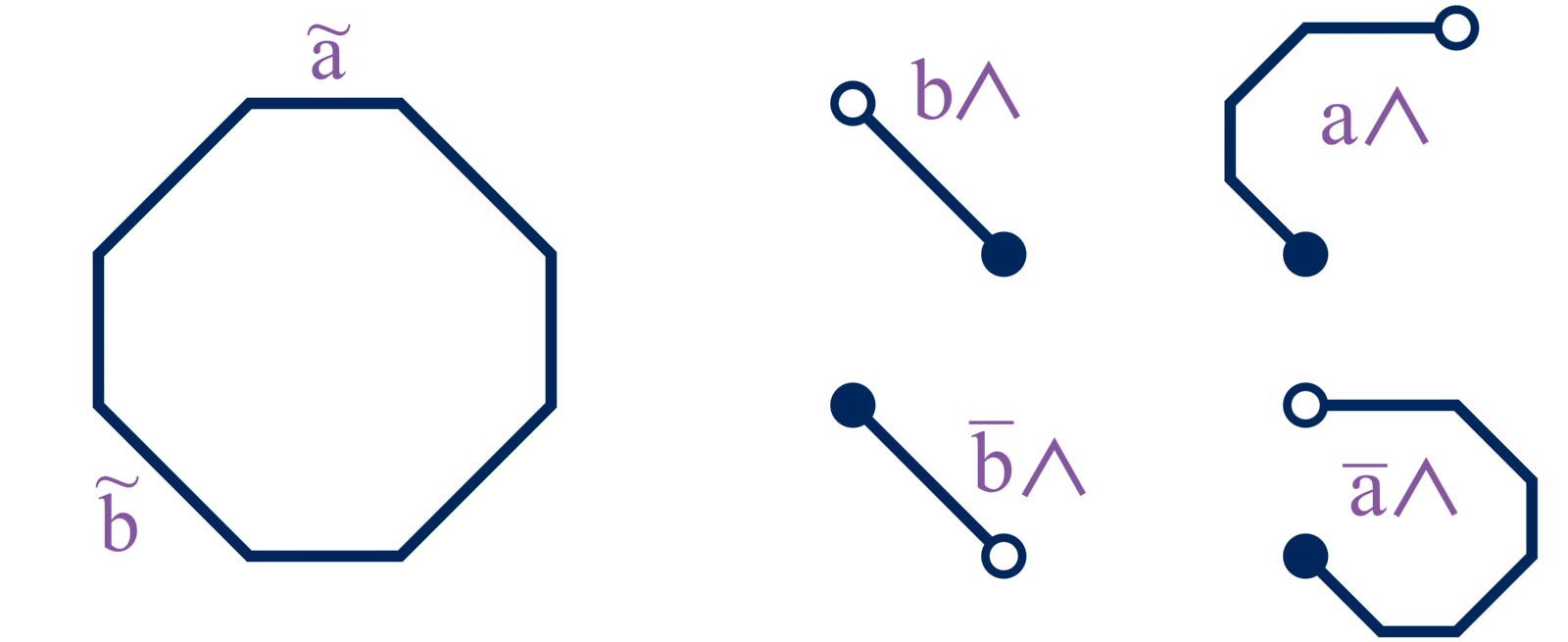}
\caption{If a path exist from $a$ to $b$ and $\br{b} \bN$ exists, then $a \bN$ exists}
\end{figure}

Suppose that stubs $b \bN$ and $\br{b} \bN$ are in $C_{\mathcal{P}}$. The set
$C^{\star}_{\mathcal{P}}$ may contain an infinite number of graphs with edges
labeled $\tilde{b}$, but all of these graphs can be reduced to $\emptyset$ with
a finite set of rules. Suppose that $G_b$ is a graph that contains an edge
labeled $\tilde{b}$. A pair of stubs $a \bN$ and $\br{a} \bN$ exists for every
labeled edge $\tilde{a}$ in $G_b$. The graph $G_b$ was constructed by gluing
path graphs or stubs. Each path graph can be reduced to a set of stubs:

\begin{figure}[H]
\centering
\includegraphics[width=4cm]{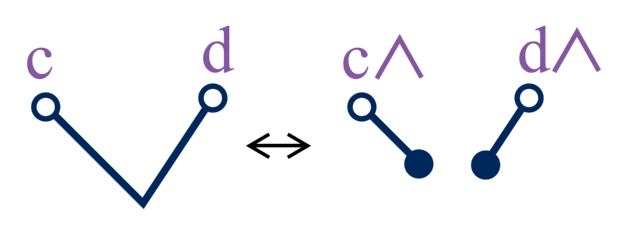}
\end{figure}

If we apply this reduction, the graph becomes a set of disconnected graphs consisting
of two opposite stubs glued together:

\begin{figure}[H]
\centering
\includegraphics[width=4cm]{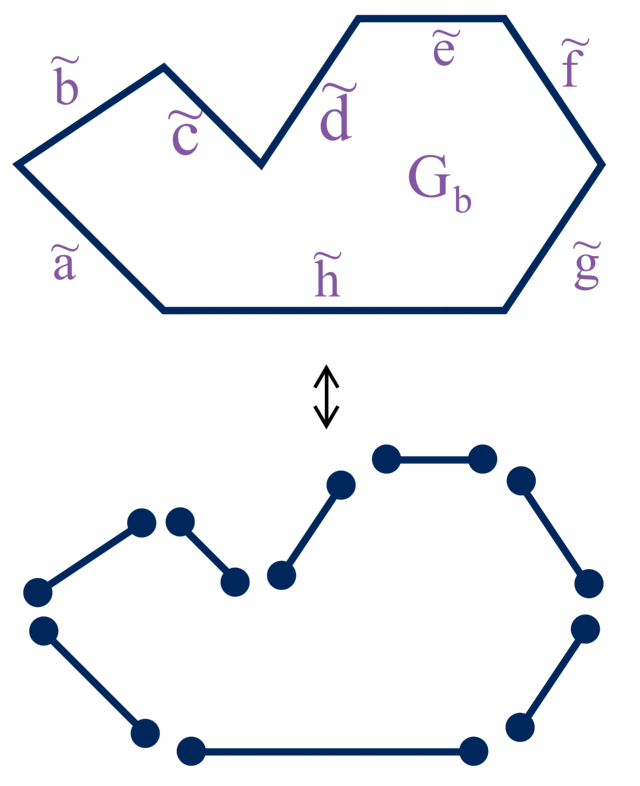}
\end{figure}

Each of graphs of two opposite stubs glued together can be reduced by a starter
rule, one rule for each edge label:

\begin{figure}[H]
\centering
\includegraphics[width=3cm]{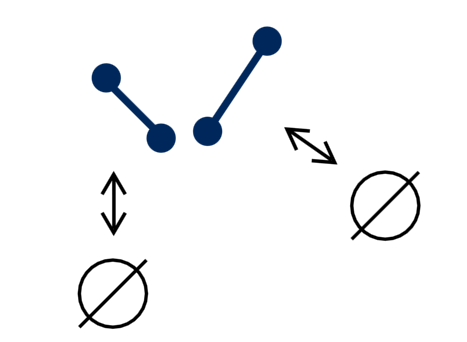}
\end{figure}

\noindent There are a finite number of edge labels, so only a finite number of starter rules are needed.

This demonstrates that if a pair of stubs $b \bN$ and $\br{b} \bN$ exists in
$C_{\mathcal{P}}$ for any half-edge $b$ in a graph $G_S$, then the graph can be reduced to
$\emptyset$ using a finite set of rules. The same finite set can be used for all
such graphs. Stubs exist for every simple path with dead ends (Fig.
\ref{stub path}). Next we consider graphs with that have no stubs
and must form a closed loop (Fig. \ref{loop path}).

\subsection{All Half-Edges Turn Strictly Upward}
\label{Strictly upward}

As shown in Section \ref{turning upward}, the half-edges in a complete loop must
all turn strictly upward, all turn strictly downward, or all turn upward and
downward. We will assume in this section that all half-edges in a complete shape
turn strictly upward. Then every half-edge $b$ in the path may only have one of
two values for its turning number either $b_i$ or $b_{i+1}$ for some number $i$.
Suppose the path has this structure:

\begin{equation}
\begin{split}
a_0 \xrightarrow{i} b_i \xrightarrow{j} b_{i+j} \xrightarrow{1-i-j} a_{1}
\end{split}
\end{equation}

Then the only possible values of $j$ are 0 or 1. If $j < 0$ that implies that
$b$ turns downward (see Eq. \ref{nonnegative}). If $j > 1$, then if we remove the
path going from $b_i \xrightarrow{j} b_{i+j}$ that implies that $a$ turns downward:

\begin{equation}
\begin{split}
a_0 \xrightarrow{i} b_i \xrightarrow{1-i-j} a_{1-j}
\end{split}
\end{equation}

Each half-edge $b$ has only two possible turning numbers $b_i$ and $b_{i+1}$ for
some $i$. We define a turned edge to be the combination of the half-edge $b$ and
the turning number $i$. There are a finite number of possible half-edge labels
and thus a finite number of possible turned edges. Any path that consists of a
finite set of turned edges can be reduced by a finite grammar $\mathcal{G}$ even
if the length of the path has no upper bound.

\begin{figure}[H]
\centering
\includegraphics[width=8cm]{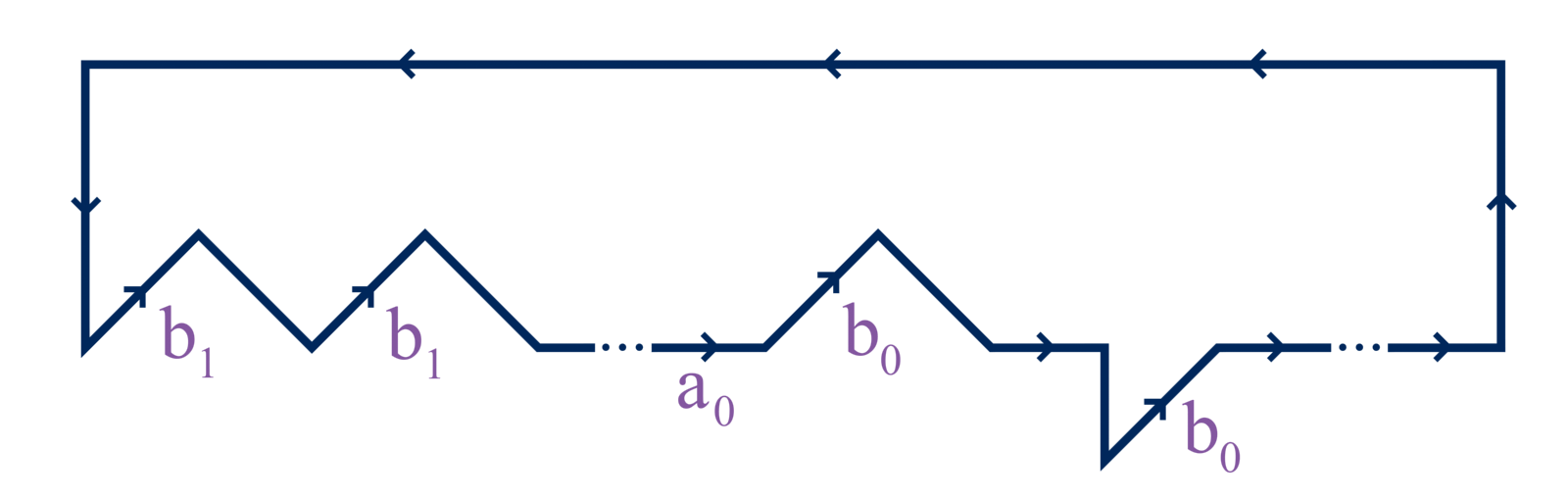}
\caption{The symbols $b_0$ and $b_1$ can repeat any number of times.}
\end{figure}

\begin{figure}[H]
\centering
\includegraphics[width=7cm]{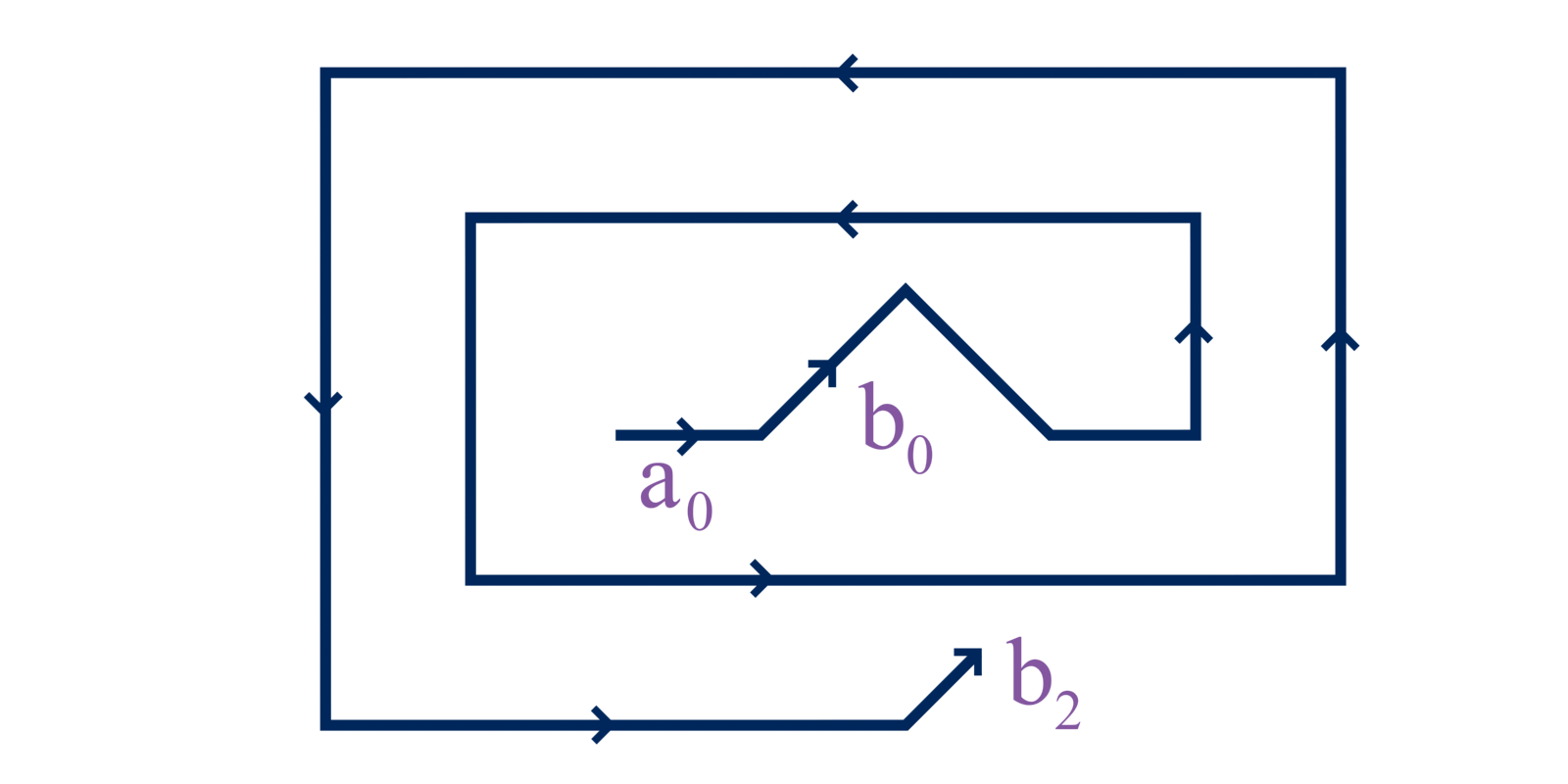}
\caption{The symbols $b_0$ and $b_2$ cannot be in the same closed loop. The loop
cannot be completed without turning downward.}
\end{figure}

We say that a path is a cycle if the turned edges repeat that is if
$b_i \xrightarrow{0} b_i$. Every cycle can be reduced by a rule. It can be
reduced to a path with just the single half-edge $b$, we can eliminate
everything in between $b_i$ and $b_i$ and replace it with a single edge $\br{b} b \bN$.

A cycle is \textit{simple} if all of its turned edges are unique and so it has no
cycles inside of it. There are a finite number of simple cycles. With $n$
distinct turned edges there are at most $n!$ simple cycles. (This is a
worst-case analysis. In practice it may be much smaller.) We
can create a rule for every simple cycle. With this finite set of rules every
path can be reduced to a path without cycles (meaning all of its turned edges
are unique). There are a finite number of paths without cycles. They can be
be reduced using a finite set of starter rules.

We have shown that any closed loop with half-edges that strictly turn upward 
can be reduced by a finite set of rules. The same argument can be made in the 
case of a path with half-edges that strictly turn downward.

\subsection{All Half-Edges Turn Upward and Downward}

The case where all half-edges turn upward and downward is different. Now it is possible
to create a closed loop where the turning numbers have no upper or lower limit.
The number of unique turned edges can be infinitely large. In this
case, we consider another type of cycle. Here we consider a cycle to be any time
the half-edge label repeats even if the turning number changes. So
$b_0 \xrightarrow{i} b_i$ would be a cycle. We begin by writing the path as a
series of simple cycles. For example, the path

\begin{equation}
ab \bN c \bN dbec \bN ce \bV fe \bV bca
\end{equation}

could be write as 

\begin{equation}
a_0 \xrightarrow{0} b_0 \xrightarrow{2} b_2  \xrightarrow{0} c_2 \xrightarrow{1} c_3 \xrightarrow{0} e_3 \xrightarrow{-1} e_2 \xrightarrow{-1} a_1
\end{equation}

Notice that besides the first and last symbol, every other symbol comes as a
pair of the form $b_i \xrightarrow{n} b_{i+n}$ describing a simple cycle. Any
simple cycle $b_0 \xrightarrow{n} b_n$ can be reduced to a series of cycles that
increment by 1:

\begin{equation}
b_0 \underbrace{\xrightarrow{1} b_1 \xrightarrow{1} \ldots \xrightarrow{1} b_n}_{\text{$n$ times}}
\end{equation}

Because the number of half-edge labels is finite, the number of possible simple
cycles is also finite. So with a finite set of rules, all these cycles can all
be reduced to a series of cycles that increment or decrement by 1. The same thing can be said of the
paths between the cycles. Any path $b_i \xrightarrow{n} c_{i+n}$ can be reduced
to the paths

\begin{equation}
b_i \underbrace{\xrightarrow{1} b_{i+1} \xrightarrow{1} \ldots \xrightarrow{1} b_{i+n}}_{\text{$n$ times}} \xrightarrow{0} c_{i+n}
\end{equation}

We can reduce every possible path to a series of cycles that only increment or
decrement by 1. Consider the case where a set of incrementing cycles is followed
by a set of decrementing cycles:

\begin{equation}
\label{incrementing}
b_{i-1} \xrightarrow{1} b_i \xrightarrow{1} b_{i+1} \xrightarrow{0} c_{i+1} \xrightarrow{-1} c_{i} \xrightarrow{-1} c_{i-1} 
\end{equation}

\noindent we can reduce the path $b_i \xrightarrow{1} b_{i+1} \xrightarrow{0} c_{i+1} \xrightarrow{-1} c_{i}$
to $b_i \xrightarrow{0} c_{i}$. If we apply this reduction twice to the above
path, it simplifies to $b_{i-1} \xrightarrow{0} c_{i-1}$. Similarly, we can reduce the
path $b_i \xrightarrow{-1} b_{i-1} \xrightarrow{0} c_{i-1} \xrightarrow{1} c_{i}$
to $b_i \xrightarrow{0} c_{i}$. We can reduce these paths using a finite set of
rules. With a finite set of rules, we can make all incrementing and decrementing
cycles cancel each other out. If we continue to cancel them out eventually we are left
with a path that contains only incrementing cycles or only decrementing cycles.

This means that the turning numbers must either monotonically increase or 
monotonically decrease. In this case, if we start the path at $a_0$, the 
turning number can never exceed 1 or go below $-1$ if the path is part of a 
closed loop. It cannot do this because the path must end at either $a_1$ 
or $a_{-1}$. We cannot get back to those values if we monotonically increase 
or decrease past them. The result, after all the reductions we applied, is that
the path now can only contain a finite number of unique turned edges. The path
started initially with an unbounded number of turned edges, but by applying these rules we have reduced 
it to a finite number.

With a finite number of unique turned edges we can apply the same argument we
in Section \ref{Strictly upward} to show that all such paths can be reduced
using a finite set of rules.

\section{Passing Through}

In the normal method, edges are not allowed to intersect. However, in some 
cases, it would be beneficial to allow the edges to intersect. It can allow 
us to solve some problems with a much simpler graph grammar. For instance, 
consider the example shape below:

\begin{figure}[H]
\centering
\includegraphics[width=8cm]{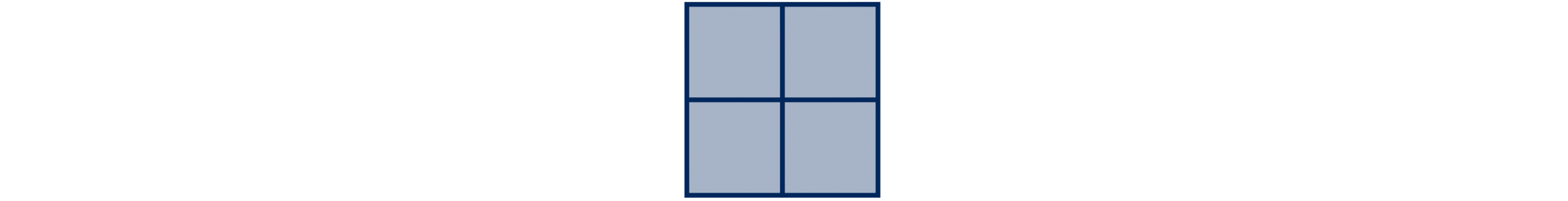}
\end{figure}

Here are some shapes that are locally similar to this example:

\begin{figure}[H]
\centering
\includegraphics[width=8cm]{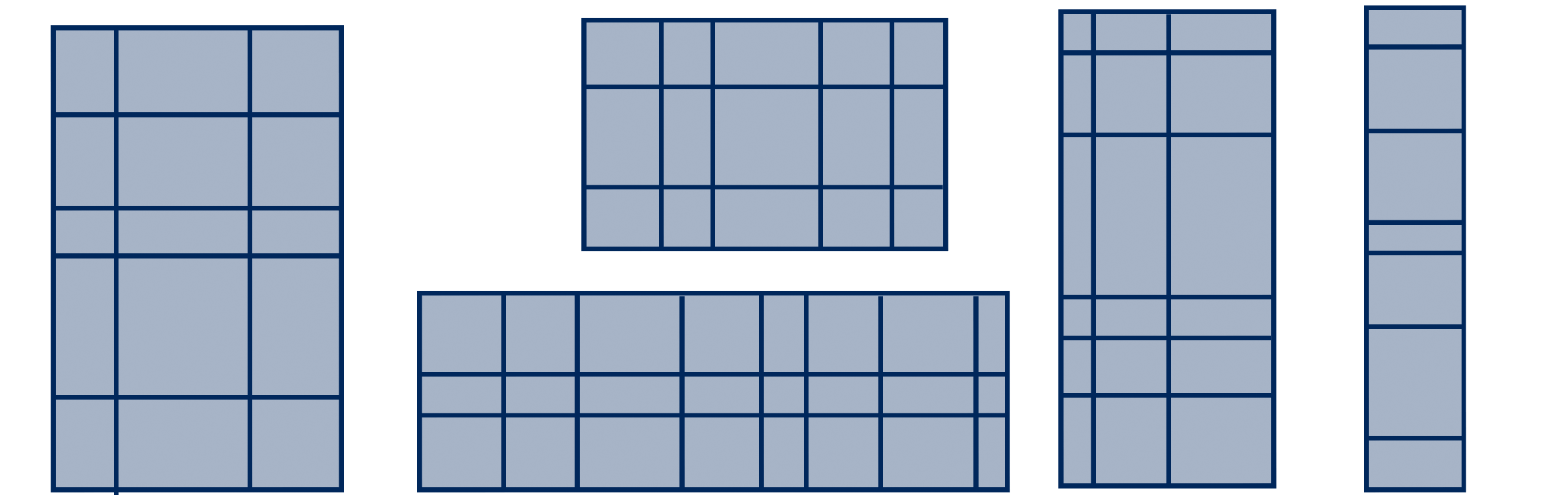}
\end{figure}

The result is always a rectangular grid with varying number of horizontal and
vertical grid lines. Any number of grid lines is allowed. We can create graph
grammar rules that add a horizontal line or a vertical line to the grid:

\begin{figure}[H]
\centering
\includegraphics[width=8cm]{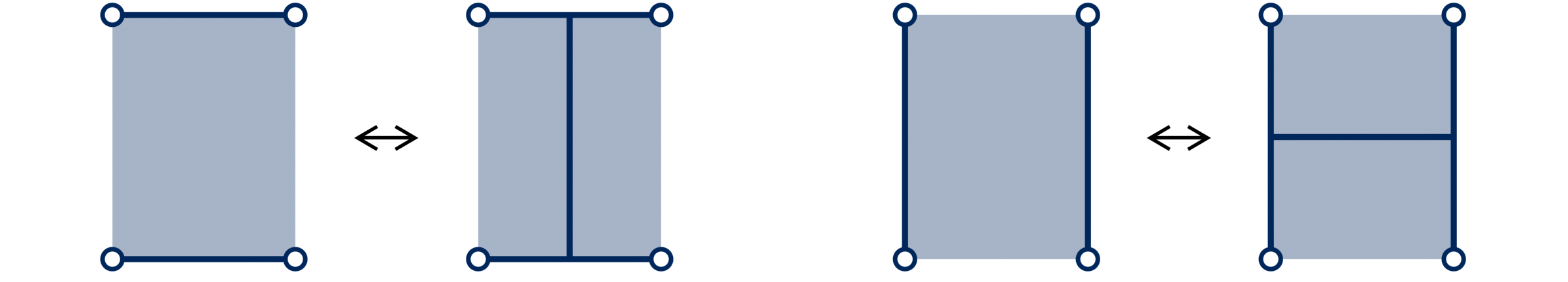}
\end{figure}

Unfortunately, the above rule for adding a vertical line can only be applied if
the grid has no horizontal lines. One could imagine generalizing the above rule
to add a vertical line, no matter how many horizontal lines there are:

\begin{figure}[H]
\centering
\includegraphics[width=8cm]{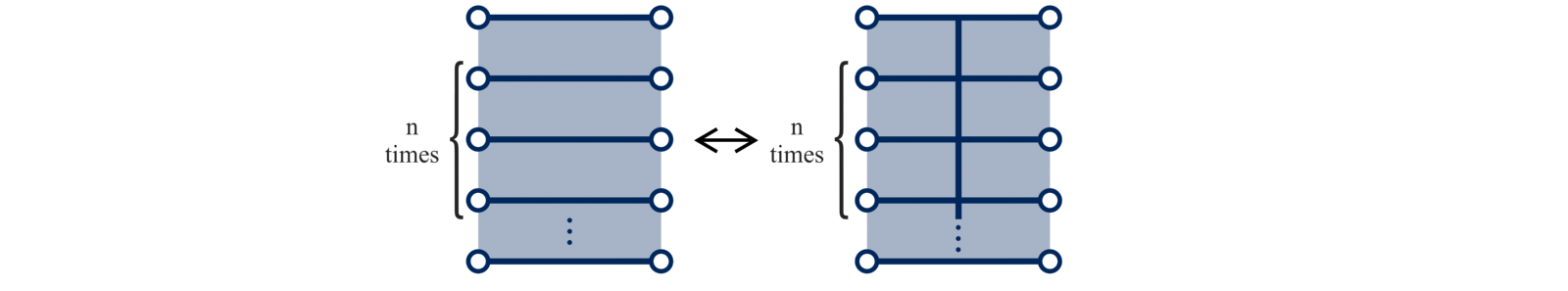}
\end{figure}

This is essentially a parameterized rule. In a some sense, it describes 
multiple rules depending on the value of the parameter $n$. Here is another 
example shapes with a set of parameterized rules that solve the problem for 
that shape:

\begin{figure}[H]
\centering
\includegraphics[width=8cm]{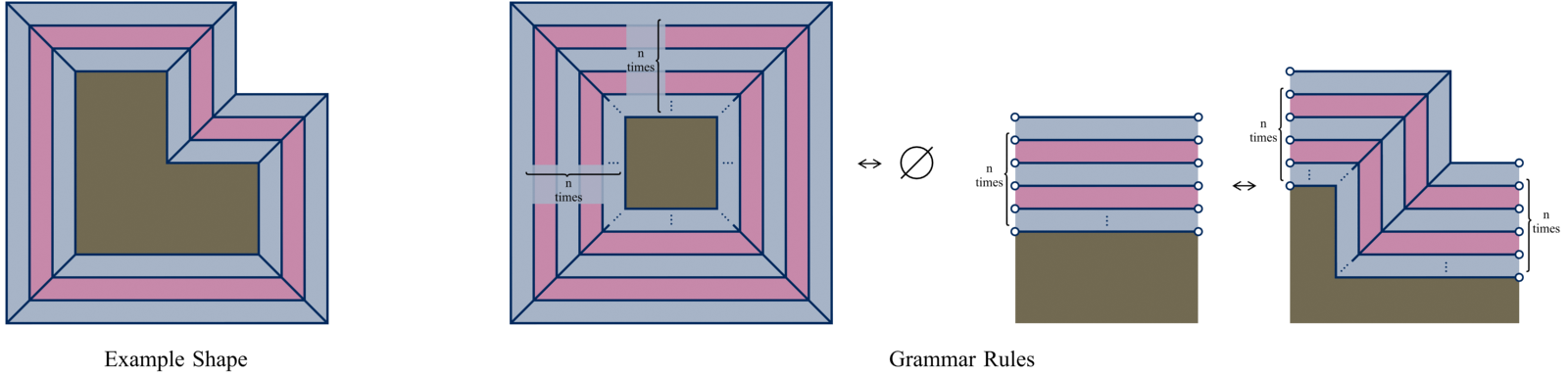}
\end{figure}

The approach of using parameterized rules has potential, but one would need a 
clear procedural for finding these rules. Another slightly different 
approach is based on a \textit{tileable} graph. Tileable graphs has 
translational symmetry both vertically and horizontally:

\begin{figure}[H]
\centering
\includegraphics[width=8cm]{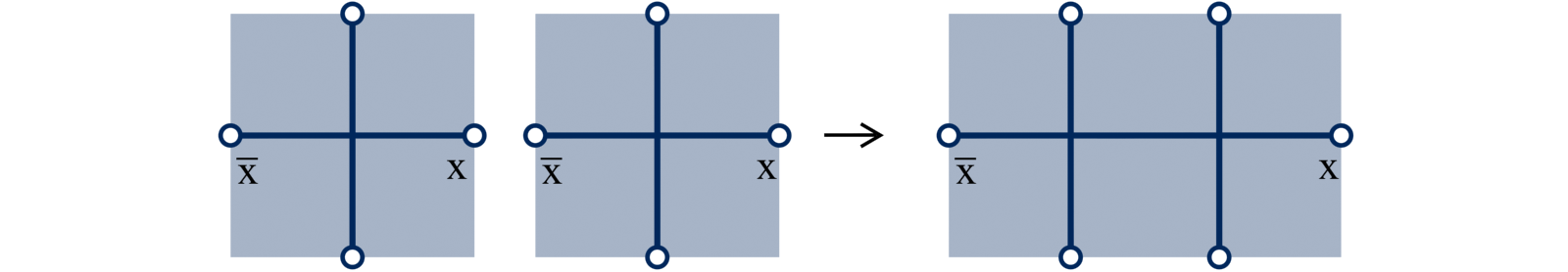}
\end{figure}

It can be glued to itself both horizontally and vertically. And it can be tiled
any number of times in either direction.

One possible approach would be to allow two edges to intersect each other if
there exists a tileable graph with those edges. There is no harm in allowing such
intersections. The intersection produces a locally similar shape. You can
imagine the edges just pass through each other or over and under one another.

This idea has not yet been implemented, but it could be useful in several 
cases. And it could be generalized to more complex intersection. Instead of 
the edges meeting at a point, they could intersect at a more complex shape 
that could be added in during the shape generation process whenever such an 
intersection occurs.

\begin{figure}[H]
\centering
\includegraphics[width=8cm]{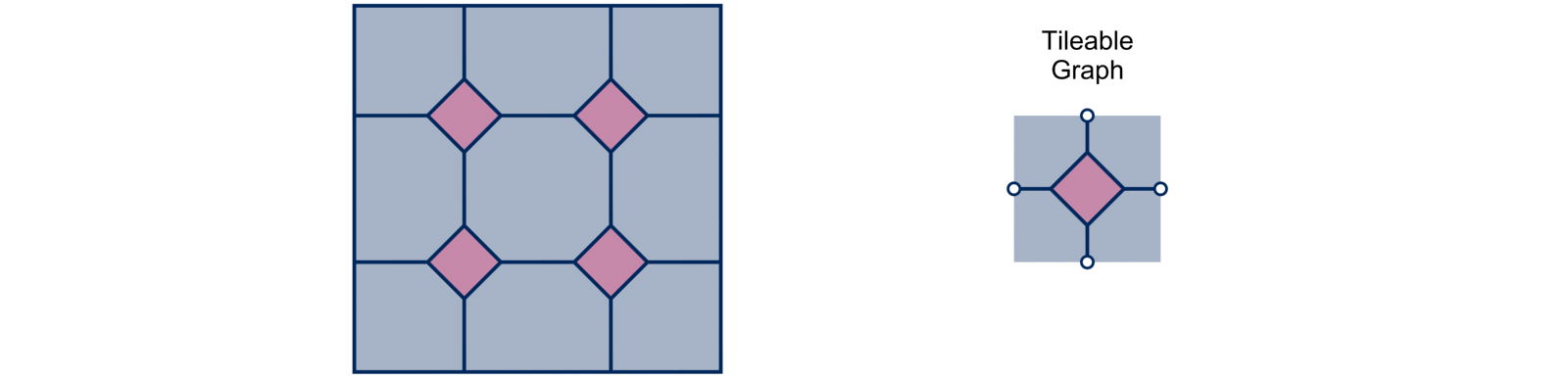}
\end{figure}

Or it could be generalized to apply to groups of edges:

\begin{figure}[H]
\centering
\includegraphics[width=8cm]{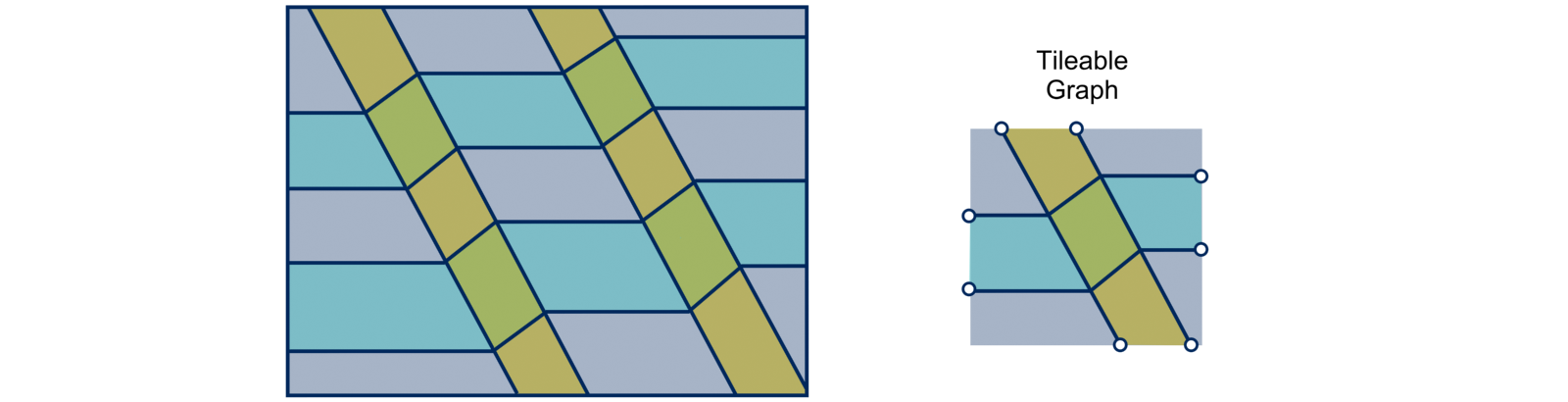}
\end{figure}

\bibliographystyle{ACM-Reference-Format}
\bibliography{companion}

\end{document}